\documentclass[10pt,aps,prl,twocolumn,superscriptaddress,floatfix,notitlepage]{revtex4-1}

\usepackage{hyperref}
\hypersetup{
    colorlinks=true,
    linkcolor=blue,
    filecolor=magenta,      
    urlcolor=blue,
}
\usepackage{amsmath,amsfonts,amssymb}
\usepackage{fancyhdr}
\usepackage{graphicx}
\usepackage{rotating}
\usepackage{braket}
\usepackage{xcolor}
\usepackage[utf8]{inputenc}

\begin{document}

\title{Hydrodynamical constraints on bubble wall velocity}
\author{Tomasz Krajewski}
\email{tkrajewski@camk.edu.pl}
 \affiliation{Nicolaus Copernicus Astronomical Center, ul. Bartycka 18, 00-716 Warsaw, Poland}
\author{Marek Lewicki}
 \email{marek.lewicki@fuw.edu.pl}
\author{Mateusz Zych}
 \email{mateusz.zych@fuw.edu.pl}
\affiliation{Faculty of Physics, University of Warsaw, ul. Pasteura 5, 02-093 Warsaw, Poland}

\begin{abstract}
Terminal velocity reached by bubble walls in first order phase transitions is an important parameter determining both primordial gravitational-wave spectrum and the production of baryon asymmetry in models of electroweak baryogenesis. We developed a numerical code to study the real-time evolution of expanding bubbles and investigate how their walls reach stationary states. Our results agree with profiles obtained within the so-called bag model with very good accuracy, however, not all such solutions are stable and realised in dynamical systems. Depending on the exact shape of the potential there is always a range of wall velocities where no steady state solutions exist. This behaviour in deflagrations was explained by hydrodynamical obstruction where solutions that would heat the plasma outside the wall above the critical temperature and cause local symmetry restoration are forbidden. For even more affected hybrid solutions causes are less straightforward, however, we provide a simple numerical fit allowing one to verify if a solution with a given velocity is allowed simply by computing the ratio of the nucleation temperature to the critical one for the potential in question.
\end{abstract}

\maketitle

%\tableofcontents

\section{Introduction}
Phase transitions are a common feature of particle physics models. If they are first order they can open a path to numerous phenomena such as the production of both the baryon asymmetry~\cite{Kuzmin:1985mm, Cohen:1993nk, Rubakov:1996vz, Morrissey:2012db} and a stochastic background of gravitational waves (GWs)~\cite{Caprini:2015zlo, Caprini:2019egz, AEDGE:2019nxb, Badurina:2021rgt}. 
Significant progress has been made recently in understanding fine details of the dynamics of such transitions necessary to describe the intricate relation between these possibilities~\cite{Cline:2020jre, Laurent:2020gpg, Dorsch:2021ubz, Dorsch:2021nje, Cline:2021iff, Cline:2021dkf, Lewicki:2021pgr, Laurent:2022jrs, Ellis:2022lft}.
Despite that, determining the bubble-wall velocity in the stationary state remains a problem. Given its impact on both the amplitude of the GW signal as well as the production of the baryon asymmetry, this issue needs to be solved in order to finally pinpoint the interplay between these two signals.

Contrary to nucleation temperature or transition strength, the wall velocity is not a straightforward consequence of the shape of the effective potential. The standard WKB method of computing the velocity involves solving a set of Boltzmann equations in the vicinity of the bubble wall in order to find the friction the plasma will enact on the expanding wall. However, the result still crucially relies on the hydrodynamical solution for the plasma profile~\cite{No:2011fi, Cline:2021iff, Lewicki:2021pgr}. 
It is a standard practice to use the plasma behaviour obtained in the bag model in these studies. The obvious drawback of this approach is that the bag equation of state (EOS) inherently neglects all knowledge of the potential except the energy difference between its minima~\cite{Espinosa:2010hh}.

In this work, we investigate the impact that features of the potential have on hydrodynamical solutions for the plasma. To this end, we perform lattice simulations tracking the real-time evolution of the scalar profiles coupled to the plasma that describe a single expanding bubble.
We use novel methods that allow us to properly resolve shocks forming in the fluid and prevent the appearance of unphysical artefacts. 
%Our method involves algebraic flux-corrected transport (FCT), described in \cite{Kuzmin:2002,Kuzmin:2003,Moller:2013,Kuzmin:2021} with an~improved version of Zalesak's limiter \cite{Zalesak:1979}.
%This allows us to study in detail how the system reaches a stationary state and compare these late-time profiles with analytical approximations.
%We take a closer look at the problem of hydrodynamical obstruction and find a large class of unstable solutions that constitute a forbidden range for bubble-wall velocities below the Jouguet velocity.  
This allows us to study for the first time in detail the evolution of the system for relatively strong transitions and fastest walls that still form heated fluid shells around bubbles. Increased temperature results in large baryon yields while large transition strength and wall velocity provide a strong GW signal making this part of the parameter space the most promising for the realisation of both~\cite{Cline:2020jre, Cline:2021dkf, Lewicki:2021pgr, Ellis:2022lft}.  
We find that the problem of hydrodynamical obstruction~\cite{Konstandin:2010dm} persists for stronger transitions and faster walls rendering a large class of solutions unstable. This constitutes a forbidden range for bubble-wall velocities particularly impacting the most promising solutions with the fastest walls which still lead to a formation of heated fluid shells surrounding the bubbles.

\section{The model: scalar field coupled to perfect fluid}
In this work, we investigate a system consisting of the scalar field $\phi$ coupled to the perfect fluid described by its temperature $T$ and local flow four-velocity $u$ \cite{Ignatius:1993qn, Kurki-Suonio:1995yaf, Kurki-Suonio:1996wfr, Hindmarsh:2013xza, Hindmarsh:2015qta}. The equation of state is given by 
\begin{align}
\epsilon(\phi,T) &= 3aT^4 + V(\phi,T) - T\frac{\partial V}{\partial T}\, ,
\label{eq:EoS1}\\
p(\phi,T) &=  aT^4 - V(\phi,T)\, ,
\label{eq:EoS2}
\end{align}
where $a=(\pi^2/90)g_*$ and $w \equiv \epsilon + p$. 
For the effective potential $V(\phi, T)$ we use a simple polynomial potential augmented with high-temperature corrections parameterized as
\begin{equation}
V(\phi,T) = \frac{1}{2}\gamma(T^2-T_0^2)\phi^2 - \frac{1}{3}\delta T\phi^3 + \frac{1}{4}\lambda\phi^4\, .
\label{VphiT}
\end{equation}
 The energy-momentum tensor of the system is a sum of energy-momentum tensors for the field and the fluid:
\begin{align}
    T^{\mu \nu}_{\textrm{field}} &= \partial^\mu \phi \partial^\nu \phi - g^{\mu \nu}\left( \frac{1}{2} \partial_\alpha \phi \partial^\alpha \phi \right)\, , \\
    T^{\mu \nu}_{\textrm{fluid}} &= w u^\mu u^\nu + g^{\mu \nu} p\, ,\label{eq:perfect_fluid_energy_momentum_tensor}
\end{align}
where $p$ is the pressure of the perfect fluid. 

The energy-momentum tensor of the system is conserved $(\nabla_\mu T^{\mu \nu} = 0 )$, however, both contributions are not conserved separately due to extra coupling term parameterised by the effective coupling of the fluid and scalar:
\begin{equation}
    \nabla_\mu T^{\mu \nu}_{\textrm{field}} = - \nabla_\mu T^{\mu \nu}_{\textrm{fluid}} = \frac{\partial V}{\partial \phi} \partial^\nu \phi + \eta u^\mu \partial_\mu \phi \partial^\nu \phi\, , \label{eq:energy_momentum_conservation}
\end{equation}
where $\eta$ is a~constant parametrizing strength of this interaction~\footnote{Assuming local thermal equilibrium and neglecting the fluid-plasma coupling $\eta$ one can obtain analytical results~\cite{Balaji:2020yrx,Ai:2021kak,Ai:2023see} for the wall velocity.}.

We use spherical coordinates in space as they capture the symmetry of a single growing bubble that we intend to simulate. The final form of equations of motion and their discretization is described in the supplemental material below.

The key parameters characterizing the transition  are the nucleation temperature $T_n$ at which at least one bubble appears per horizon and the amount of the vacuum energy released in the transition normalized to the energy of the radiation bath $\rho_r$. In the fluid approximation, it can be defined as 
\begin{equation}
\alpha = \frac{\theta_s-\theta_b}{\rho_r}\Big{|}_{T=T_n}, 
\label{eq:alpha}
\end{equation}
where $\theta$ is the trace anomaly in the symmetric (s) and broken (b) phase, given by the expression:
\begin{equation}
 \theta = \frac{1}{4}(\epsilon - 3p).
\end{equation}
Note, that this definition of the trace anomaly applied to the equation of state \eqref{eq:EoS1}-\eqref{eq:EoS2} corresponds to the standard definition of $\alpha = \frac{1}{\rho_r}\left(\Delta V - \frac{T}{4}\Delta\frac{\partial V}{\partial T}\right)$ \cite{Hindmarsh:2017gnf, LISACosmologyWorkingGroup:2022jok}.

%%%%%%%%%%%%%%%%%%%%%%%%%%%%%%%%%%%%%%%%%%%%%%%%%%%%%%%%%%%%%%%%%%%%%%%%%%%%%%%%%%%%%%%%%%%%%%%%%%%%%%%%%%%%%%%%%%%%%%%%%%
%%%%%%%%%%%%%%%%%%%%%%%%%%%%%%%%%%%%%%%%%%%%%%%%%%%%%%%%%%%%%%%%%%%%%%%%%%%%%%%%%%%%%%%%%%%%%%%%%%%%%%%%%%%%%%%%%%%%%%%%%%
%%%%%%%%%%%%%%%%%%%%%%%%%%%%%%%%%%%%%%%%%%%%%%%%%%%%%%%%%%%%%%%%%%%%%%%%%%%%%%%%%%%%%%%%%%%%%%%%%%%%%%%%%%%%%%%%%%%%%%%%%%
\section{Analytical approximation: bag model}
A simple model describing analytically many important features of the late-time evolution of our system is the bag model \cite{Espinosa:2010hh}. It neglects the scalar field approximating its profile as a step function connecting the two vacua.
The equation of state reads
\begin{align}
    \epsilon_{s} &= 3 a_s T_s^4 + \theta_{s}& \epsilon_{b} &= 3 a_b T_b^4 + \theta_{b} \\
    p_{s} &= a_s T_s^4 - \theta_{s} &
    p_{b} &= a_b T_b^4 - \theta_{b},  
\end{align}
where $\theta_s$ and $\theta_{b}$ correspond to the symmetric phase outside the bubble and broken phase inside respectively. 
Assuming that the vacuum energy vanishes in the broken phase as the field is in the global minimum of the potential we have $\theta_{b}=0$.

 Therefore the strength of the transition can be consistently defined with the equation \eqref{eq:alpha}. Assuming that the plasma is locally in equilibrium, the energy-momentum tensor can be parameterized for the perfect fluid as in eq. \eqref{eq:perfect_fluid_energy_momentum_tensor}. %:
%\begin{equation}
%    T^{\mu\nu} = w u^{\mu}u^{\nu} - g^{\mu\nu}p.
%\end{equation}
Conservation of $T^{\mu \nu}_{\textrm{fluid}}$ along the flow and its projection perpendicular to the flow respectively give
\begin{align}
    \partial_{\mu}( u^{\mu} w) - u_{\mu}\partial^{\mu}p &= 0,
    \label{eq:bag1}\\
    \bar{u}^{\nu}u^{\mu} w\partial_{\mu}u_{\nu} - \bar{u}^{\nu}\partial_{\mu}p &= 0,
    \label{eq:bag2}
\end{align}
with $\bar{u}_{\mu}u^{\mu} = 0$ and $\bar{u}^2 = -1$. As
there is no characteristic distance scale in the problem, the solution should depend only on the self-similar variable $\xi = r/t$, where $r$ denotes the distance from the center of the bubble and $t$ is the time since nucleation. Changing the variables, equations \eqref{eq:bag1} and \eqref{eq:bag2} take the form
\begin{align}
    (\xi-v)\frac{\partial_{\xi} \epsilon}{w} &= 2\frac{v}{\xi}+[1-\gamma^2v(\xi-v)]\partial_{\xi}v,\\
    (1-v\xi)\frac{\partial_{\xi} p}{w} & = \gamma^2(\xi-v)\partial_{\xi}v
\end{align}
and using the definition of the speed of sound in the plasma $c_s^2 \equiv \frac{\textrm{d}p}{dT}/\frac{\textrm{d}\epsilon}{dT}$ can be combined into the single equation describing the plasma velocity profile $v(\xi)$ in the frame of the bubble center
\begin{equation} 
2\frac{v}{\xi} =\gamma^2 (1 - v \xi )\left[\frac{\mu^2}{c_s^2}-1\right]\partial_{\xi}v,
\label{eq:bag_central}
\end{equation}
with $\mu = \frac{\xi-v}{1-\xi v}$ denoting the Lorentz-transformed fluid velocity. Solutions of the equation \eqref{eq:bag_central} in general depend only on the transition strength $\alpha$ and bubble-wall velocity in the stationary state $\xi_w$. In a similar way, analytical profiles for the enthalpy $w$, temperature $T$ and other thermodynamical quantities can be obtained. Later we will refer to them to compare the results of our simulations with the analytical solutions.  Detailed derivations are described in \cite{Espinosa:2010hh,Hindmarsh:2019phv,Ellis:2020awk,Lewicki:2021pgr}. 
In general, there exist three types of the bubble-wall profiles:
\begin{enumerate}
    \item \textbf{Deflagrations} are the solutions with subsonic bubble-wall velocity $\xi_w$. In such a case, the expanding bubble pushes the plasma in front of it, while behind the bubble wall plasma remains at rest. Typically, value of $v$ decreases with $\xi$ in the range $[\xi_w, c_s]$ and vanishes for $\xi>c_s$. Therefore a shock front at $\xi = c_s$ may appear if the transition is strong enough.
    \item \textbf{Detonations} are supersonic solutions, for which bubble-wall velocity exceeds the Jouget velocity $c_J$ (see Eq.~\eqref{eq:vJ}). In this type of profile, the wall hits plasma which remains at rest in front of the bubble. As fluid enters the broken phase, it is accelerated with its velocity decreasing smoothly and reaching zero at $\xi = c_s$.
    \item \textbf{Hybrids} are combinations of the two types mentioned above. They are realised for $\xi_w \in [c_s, c_J]$ and possess features of deflagrations (shock front in front of the wall) and detonations (non-zero plasma velocity behind the wall known as a~rarefaction wave). 
\end{enumerate}

 The Jouget velocity $c_J$ at which the shell around the bubble disappears and the solution shifts from hybrid to detonation is given by Chapman-Jouguet condition~\cite{Steinhardt:1981ct, Lewicki:2021pgr}
\begin{equation}\label{eq:vJ}
    c_J = \frac{1}{\sqrt{3}}\frac{1+\sqrt{3\alpha^2 + 2\alpha}}{1+\alpha}.
\end{equation}

This is a crucial threshold as in particle physics models the friction typically grows with the temperature. Thus, if for a given potential the strength is large enough for the wall to accelerate above this velocity, the disappearance of the heated fluid shell around the bubble also lowers the friction substantially. As a result, in simple extensions of the Standard Model, one does not expect to find detonations with wall velocity significantly below the speed of light. While such very fast walls are optimal for a strong GW signal they do not generate a significant baryon asymmetry. In fact, it is the velocity just below $c_J$ that prove to produce the largest baryon yield due to increased temperature in the plasma surrounding the bubble~\cite{Cline:2021dkf, Lewicki:2021pgr, Laurent:2022jrs, Ellis:2022lft}. However, for the purpose of our scans, the non-equilibrium part of the friction is a free parameter and we adjust it to obtain all stable solutions including detonations.

%%%%%%%%%%%%%%%%%%%%%%%%%%%%%%%%%%%%%%%%%%%%%%%%%%%%%%%
%%%%%%%%%%%%%%%%%%%%%%%%%%%%%%%%%%%%%%%%%%%%%%%%%%%%%%%
%%%%%%%%%%%%%%%%%%%%%%%%%%%%%%%%%%%%%%%%%%%%%%%%%%%%%%%
%%%%%%%%%%%%%%%%%%%%%%%%%%%%%%%%%%%%%%%%%%%%%%%%%%%%%%%

\section{Results from numerical simulations} \label{sec:results}

In this section, we will discuss the results of our numerical simulations.
Every simulation is performed on the lattice with spacing $\delta r = 0.01$ GeV$^{-1}$ and time step $\delta t = 0.001$ GeV$^{-1}$. The time duration of the evolution is large enough to asymptotically achieve stationary states and is set to $t_{max} = 120$ GeV$^{-1}$. Similarly, the physical size of the lattice is fixed as $R = c t_{max}$ which is large enough to prevent reaching the boundaries by the bubbles, since they expand subluminally.

We initialize each simulation with the recently nucleated bubble, fixing the field configuration to the critical profile and setting $T=T_n$ and $v=0$ everywhere. The procedure for determining nucleation temperature $T_n$ and the critical profile is described in supplemental material.

We have begun with the validation of our method on two benchmark points studied in the existing literature. We also compared our stationary states and those predicted by the bag model described in the previous section finding very good agreement. Results and details of both comparisons can be found in the supplemental material.

%%%%%%%%%%%%%%%%%%%%%%%%%%%%%%%%%%%%%%%%%%%%%%%%%%%%%%%
%%%%%%%%%%%%%%%%%%%%%%%%%%%%%%%%%%%%%%%%%%%%%%%%%%%%%%%
%%%%%%%%%%%%%%%%%%%%%%%%%%%%%%%%%%%%%%%%%%%%%%%%%%%%%%%
%%%%%%%%%%%%%%%%%%%%%%%%%%%%%%%%%%%%%%%%%%%%%%%%%%%%%%%
%\subsection{Dependence on the temperature}

The exact relation between the wall velocity $\xi_v$ and the friction parameter $\eta$ is not unique and depends not only on the strength of the transition $\alpha$, but also on other parameters defining the scalar potential. We start with the dependence of the results on the vacuum expectation value of the scalar field. We verified that the friction scales proportionally to the inverse of field value in the true vacuum. This fully determines the position of the gap in terms of friction parameter and to provide general results we will use the normalised value of the friction $v_0 \eta$ already including this scaling.

Next, we move on to the much more interesting dependence of the results on the nucleation temperature  $T_n$. We study a set of different potentials for which the transition strength $\alpha$ and the critical temperature $T_c$, at which the two minima in the potential are degenerate, are both fixed. The other parameters in the potential are chosen such that they predict a set of different nucleation temperatures.
The value of the friction parameter depends on the field content of the model and as a result, we keep it as a free parameter. We logarithmically vary it in the range $\eta/T_c\in[0.01, 1]$, independently checking around 75 values for every scalar potential which is enough to map all the allowed classes of solutions in each case.

\begin{figure}[ht]
    \centering
    \includegraphics[scale=1]{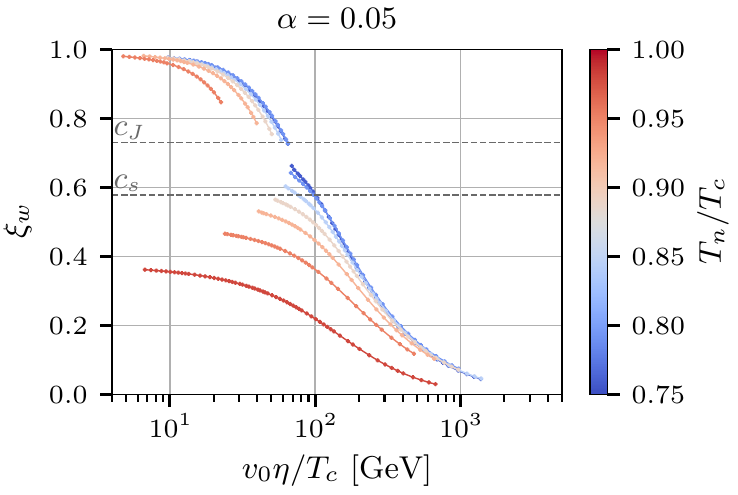}
    \includegraphics[scale=1]{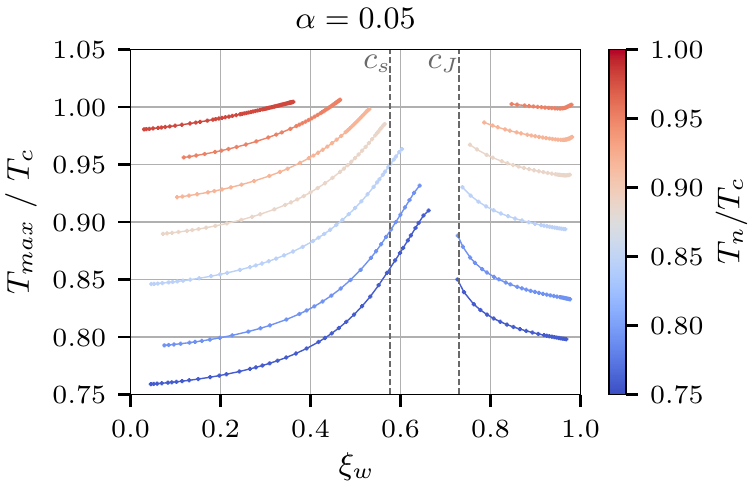}
    \caption{\it{Relation between the friction parameter $\eta$ and bubble-wall velocity in the stationary state $\xi_w$ (upper panel) and the maximum of the plasma temperature along the profile (lower panel). Potentials are chosen such that the strength of the transition is fixed to $\alpha =0.05$. The value of the nucleation temperature $T_n$ is encoded with the colour.}}
    \label{fig:summary_Tn1}
\end{figure}

\begin{figure}[ht]
    \centering
    \includegraphics[scale=1]{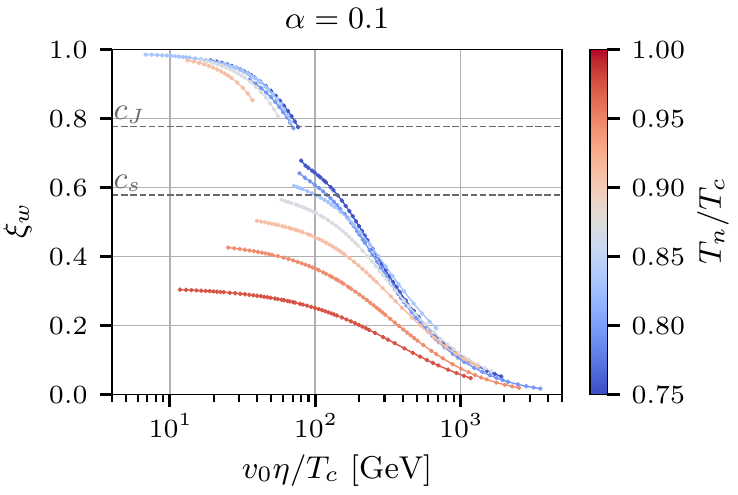}
    \includegraphics[scale=1]{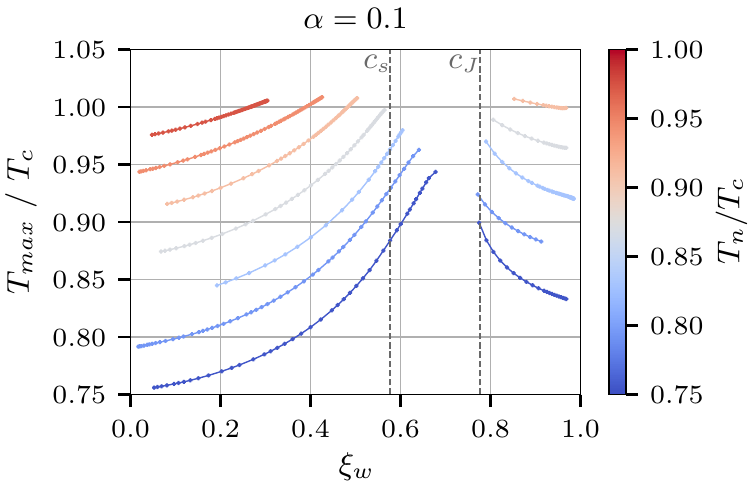}
        \caption{\it{Relation between the friction parameter $\eta$ and bubble-wall velocity in the stationary state $\xi_w$ (upper panel) and the maximum of the plasma temperature along the profile (lower panel). Potentials are chosen such that the strength of the transition is fixed to $\alpha =0.1$. The value of the nucleation temperature $T_n$ is encoded with the colour.}}
    \label{fig:summary_Tn2}
\end{figure}

%The exact relation between the wall velocity $\xi_v$ and the friction parameter $\eta$ is not unique and depends not only on the strength of the transition $\alpha$, but also on other parameters defining the scalar potential. To illustrate this, we study a set of different potentials for which $\alpha$ and $T_c$ are fixed, however, the parameters in the potential are chosen such that they predict different nucleation temperatures $T_n$. 
Relations between the friction $\eta$ and wall velocity $\xi_w$ in a range of $T_n$ are shown the upper panels of Fig \ref{fig:summary_Tn1} and~\ref{fig:summary_Tn2} for $\alpha =0.05$ and $\alpha = 0.1$ respectively.

There we can see that higher nucleation temperatures lead to a wider velocity gap, while for lower temperatures, almost the entire range of wall velocities can be covered. Note that nucleation temperatures very close to the critical temperature limit the bubble wall velocity for both deflagrations and hybrids so the speed of sound is never reached.

This dependence is made clearer in the lower panels of Fig \ref{fig:summary_Tn1} and \ref{fig:summary_Tn2}, where we show values of temperature at the peak of the bubble-wall profile for different nucleation temperatures and values of $\alpha$. As we see in general it is not possible to find a stationary state if the temperature in the profile significantly exceeds the critical temperature. This is an important condition for the part of the velocity gap below $c_s$ and indeed this hydrodynamical obstruction was already proposed in the small velocity limit~\cite{Konstandin:2010dm}, where the authors derived an approximation for the maximal subsonic wall-velocity. 
Our results agree roughly with this limit when the nucleation temperature is very close to the critical one. However, we found that a similar behaviour continues for much lower temperatures and also supersonic solutions are affected. The mechanism itself in those cases becomes less straightforward as the instability sets in when the temperature reached within the shells is significantly below the critical one.

\begin{figure}[!h]
\centering
\includegraphics[scale=1]{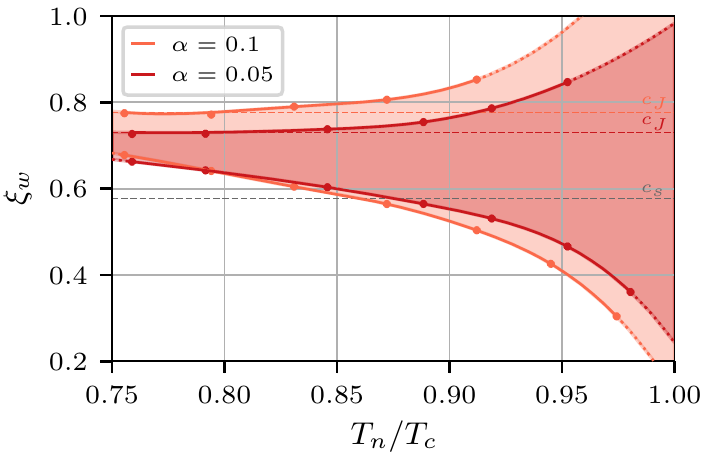}
\caption{\it Forbidden wall velocities (shaded regions) for two values of $\alpha=0.1$ (orange) and $\alpha=0.05$ (red). Dashed lines indicate the Jouguet velocities $c_J$ (see Eq.~\eqref{eq:vJ}) and the speed of sound $c_s$. At high temperatures, hydrodynamic obstruction limits the velocities of both detonation and deflagration solutions. At lower temperatures detonations are realised above the Jouguet velocity as expected, however, we find the obstruction limiting the maximal velocity of hybrids persists resulting in a range of hydrodynamical solutions that are not realised.
High-temperature part of the limit relies on extrapolation (dotted lines) and should be treated as a qualitative trend.
}
\label{fig:regions}
\end{figure}

Our results show that a range of wall velocities is forbidden and the width of the gap depends on the relation between the critical and nucleation temperatures. The forbidden region is shown in Fig.~\ref{fig:regions} for the relatively large values of the transition strength which we studied. 
The upper limit showing the slowest possible detonations is explained through the Jouguet velocity at low temperatures and excessive heating when approaching the critical temperature. The lower limit corresponding to the fastest solutions predicting a heated fluid shell around the bubble is our main result. At high temperatures, it also corresponds to heating of the shell above the critical temperature yet it persists into supersonic solutions where the maximal temperature does not reach the critical one.  

\begin{figure}[ht]
\centering
\includegraphics[scale=1]{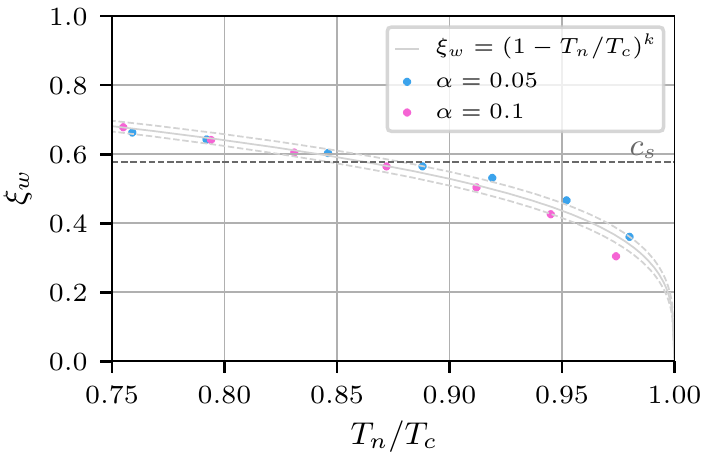}
\caption{\it The maximal wall velocity on the deflagration/hybrid branch computed in numerical simulations as a function of the nucleation temperature $T_n$ together with the fit from Eq.~\eqref{eq:vwfit} (solid line) and its variation within $3\sigma$ of the best-fit parameters (dashed lines).
}
\label{fig:fit}
\end{figure}

Fig.~\ref{fig:fit} shows the maximal wall velocity reached by the deflagration/hybrid solutions as a function of the nucleation temperature for different transition strengths. Given that in limit of relatively large strength the result for hybrids does not depend significantly on $\alpha$ we found a simple fit
\begin{align}\label{eq:vwfit}
 \xi_w^{max} = \left(1-\frac{T_n}{T_c}\right)^k &&\textrm{with} && k =0.2768 \pm 0.0055,
\end{align}
also shown in the plot, which can be used as a rough approximation for the upper bound for wall velocities. %It has a similar origin as the relation proposed in \cite{Konstandin:2010dm}, but is augmented with additional suppression that we observed for low temperatures, described by the power $k$.

Direct verification of this mechanism in particle physics models is beyond the scope of this work, however, to assess its impact we checked our criterion against the points from scans of the Standard Model plus a neutral singlet presented in ref.~\cite{Ellis:2022lft}. We found that nearly all wall velocities predicted there will be lowered as the walls cease to accelerate due to the obstruction before achieving the velocities predicted by the WKB approximation. We discuss the details of this comparison in the last section of the supplemental material.
%%%%%%%%%%%%%%%%%%%%%%%%%%%%%%%%%%%%%%%%%%%%%%%%%%%%%%%%%
%%%%%%%%%%%%%%%%%%%%%%%%%%%%%%%%%%%%%%%%%%%%%%%%%%%%%%%%%

\section{Summary} \label{sec:conclusion}

We investigate the fluid solutions realised in the presence of growing bubbles in cosmological first order phase transitions. We use numerical lattice simulations using spherical symmetry of the system and compare results to the well known analytical solutions.

We found good agreement between the analytical profiles and our numerical results whenever the latter exist. Our key result, however, is that the hydrodynamical obstruction preventing the realisation of fast hybrids is very generic. In fact, we always find some solutions to be forbidden and the gap in solutions becomes wider as the temperature at which bubbles nucleate predicted by the potential is closer to the critical temperature at which the minima in the potential are degenerate. In extreme cases where the temperatures are very close, no hybrid solutions are realised and as the friction drops the allowed solutions jump from subsonic deflagrations straight to detonations. The mechanism behind the obstruction is well understood in the case of deflagrations where the temperature profiles in the gap that are not realised would simply heat the plasma above the critical temperature and reverse the transition. In the case of hybrids the mechanism is more complicated and even solutions that do not reheat to such dangerous levels are not realised.

While the effect is yet to be confirmed directly in particular models we expect it to be general. Our calculations were performed for a simple toy potential, however, we express them in terms of general characteristics shared by all models predicting a first order transition.

The existence of the velocity gap will have a crucial impact on predictions of models realising electroweak baryogenesis. This is due to the fact that the fastest walls that did not accelerate enough to become detonations are the ones most likely to be affected and the effect would persist even in low temperatures. Such solutions were recently shown to predict the largest baryon yields. Thus our results are likely to exclude parts of the parameter space of models most promising for electroweak baryogenesis and impact their viability as solutions to the problem of baryon asymmetry.

\begin{acknowledgments}
\section*{Acknowledgments}
The authors would like to thank Jos\'e Miguel No for fruitful discussions. This work was supported by the Polish National Agency for Academic Exchange within Polish Returns Programme under agreement PPN/PPO/2020/1/00013/U/00001 and the Polish National Science Center grant 2018/31/D/ST2/02048. T.K. was supported by grant 2019/32/C/ST2/00248 from the Polish National Science Centre. During the completion of this work, T.K. was supported by grant 2019/33/B/ST9/01564 from the Polish National Science Centre. T.K. acknowledges the hospitality of Rudolf Peierls Centre for Theoretical Physics at Oxford University, where parts of this work have been done.
\end{acknowledgments}

\newpage
\appendix

\section{SUPPLEMENTAL MATERIAL}\label{sec:appendix}
In this supplemental material, we discuss the technical details of numerical calculations leading to our main result. We begin with the exact form of the equations of motion including the symmetry we assumed and methods we use to implement and solve them on a numerical lattice. Next, we show the results we obtained for benchmark points discussed in the past. Finally, we give simple analytical approximations used to scan the parameter space of couplings in the potential to find relevant examples. 

\section{Equations of motion and their discretization\label{sec:discretisation}}
The energy-momentum tensor of the system is conserved $(\nabla_\mu T^{\mu \nu} = 0 )$, however, both contributions are not conserved separately due to extra coupling term parameterised by the effective coupling of the fluid and scalar:
\begin{equation}
    \nabla_\mu T^{\mu \nu}_{\textrm{field}} = - \nabla_\mu T^{\mu \nu}_{\textrm{fluid}} = \frac{\partial V}{\partial \phi} \partial^\nu \phi + \eta u^\mu \partial_\mu \phi \partial^\nu \phi\, ,
\end{equation}
where $\eta$ is a~constant parametrizing strength of this interaction.
We use spherical coordinates in space as they capture the symmetry of a single growing bubble that we intend to simulate~\footnote{We neglect the possible deviations to spherical symmetry coming from instabilities that have been suggested in planar wall propagation~\cite{Kamionkowski:1992dc}.}.
The line element $ds$ in flat space-time takes the following form in these coordinates:
\begin{equation}
    ds^2 = - dt^2 + dr^2 + r^2 \left(d\theta^2 + \sin^2{\theta} d \varphi^2 \right)\, .
\end{equation}
The left-hand side of the equation \eqref{eq:energy_momentum_conservation} contains wave equation in spherical coordinates and leads to the equation of motion
\begin{equation}
    -\partial_t^2\phi + \frac{1}{r^2}\partial_r(r^2\partial_r\phi) -
\frac{\partial V}{\partial\phi}
  =  \eta\gamma(\partial_t\phi + v\partial_r\phi)\, \label{eq:nonlinear_wave_equation}.
\end{equation}
Due to the spherical symmetry of our problem, the four-velocity of the perfect fluid takes the form $ u = (\gamma, \gamma v, 0, 0)^T$ with $\gamma:= (1 - v^2)^{-1}$. We will determine the equations governing the evolution of two parameters $v$ and $p$ considering temporal ($\nu = 0$) and radial component ($\nu=1$) of Eq.~\eqref{eq:energy_momentum_conservation}. %:
%\begin{subequations}
%\begin{align}
%    \nabla_\mu T^{\mu 0}_{\textrm{fluid}} &=  \partial_t (w \gamma^2 - p) + \frac{1}{r^2} \partial_r (r^2 w \gamma^2 v)\, , \\
%    \nabla_\mu T^{\mu 1}_{\textrm{fluid}} &= \partial_t (w \gamma^2 v) + \frac{1}{r^2} \partial_r \left(r^2 w \gamma^2 v^2\right) + \partial_r p\, .
%\end{align}
%\end{subequations}
Introducing new variables $Z:=w\gamma^2v$ and $\tau:=w\gamma^2 - p$ we get
\begin{subequations}\label{eq:plasma_equations_of_motion}
\begin{align}
&\begin{aligned}
\nabla_\mu T^{\mu 0}_{\textrm{fluid}} &= \partial_t \tau + \frac{1}{r^2} \partial_r (r^2 (\tau + p) v)\\
&= \frac{\partial V}{\partial \phi} \partial_t \phi + \eta \gamma (\partial_t \phi + v \partial_r \phi) \partial_t \phi\, ,\\
\end{aligned}\\
&\begin{aligned}
\nabla_\mu T^{\mu 1}_{\textrm{fluid}} &= \partial_t Z + \frac{1}{r^2} \partial_r \left(r^2 Zv \right) + \partial_r p\\
&= -\frac{\partial V}{\partial \phi} \partial_r \phi  -\eta \gamma (\partial_t \phi + v \partial_r \phi) \partial_r \phi\, .
\end{aligned}
\end{align}
\end{subequations}
The final step needed to evolve the system numerically is to discretize and solve the equations presented above. 
In order to obtain a numerical approximation for solutions of equations \eqref{eq:nonlinear_wave_equation} and \eqref{eq:plasma_equations_of_motion} we use the finite element method, both in time and space.

To discretise in space we used the discontinuous Galerkin method. Our elements are just intervals of length $\delta r$ in the computational domain $[0, R]$. We used the value of $R$ large enough to guarantee that the wall of the bubble is far from $r=R$ during the whole simulation, thus the choice of the boundary condition at this point does not influence the results.

Wave equation \eqref{eq:nonlinear_wave_equation} describing the evolution of the scalar field is treated with a mixed scheme, i.e. we introduced auxiliary variable $\psi = \partial_r \phi$ which is interpolated using discontinuous piece-wise linear interpolation functions. Using the generalised trapezoid rule as numerical quadrature we obtained a second order scheme which is a generalization of the central finite difference scheme of the second order in Cartesian coordinates. 

In the center of the bubble ($r=0$) we assumed the Neumann boundary condition for field $\phi$ which in the mixed formulation is just $\left.\psi\right|_{r=0} = 0$. At the far edge of the computational lattice ($r=R$), we assumed the Dirichlet boundary condition setting the field value to the location of the false minimum. 

In order to discretize the equation of motion of the field in time we used the discontinuous Galerkin method \cite{Tang:2012, Gagarina:2013, Zhao:2014, Campos:2014, Zhao:2014, Oberblobaum:2014, Gagarina:2016, Muehlebach:2016, Ober-Blobaum:2016}. The discontinuous piece-wise linear interpolation functions for $\phi$ and right-discontinuous linear interpolation for time derivative $\dot{\phi}$ result in a scheme mimicking the well-known position version of Str\"omer-Verlet scheme.

Deriving a numerical scheme for equations describing the evolution of plasma is somewhat more involved. We base our method on algebraic flux-corrected transport (FCT) proposed in \cite{Kuzmin:2002, Kuzmin:2003, Moller:2013, Kuzmin:2021}. Since fluxes in Eq. \eqref{eq:plasma_equations_of_motion} are determined in terms of both conserved and so-called primitive variables $v$, $T$ (and derived from them $p$) one has to determine primitive ones from $\phi$, $\tau$ and $Z$ which are evolved in the code. To do so we combine equations of state to find
 \begin{equation}
     \tau + p(\phi, T) - \frac{1}{2}\left(w(\phi, T) +\sqrt{w(\phi,T)^2+ 4Z^2} \right) = 0
 \end{equation}
which we solve using the Raphson-Newton method for the value of the temperature $T$. Then, $w$ and $p$ can be directly computed and the velocity $v$ can be simply computed by inverting the definition of $Z$.

In order to derive the high-order (in our case second order) scheme for the FCT procedure we used the local discontinuous Galerkin method with piece-wise constant interpolation functions for conserved quantities $Z, \tau$, thus our scheme is similar to a finite volume method. To discretize the high-order scheme in time we use the midpoint method which can be derived as the discontinuous Galerkin method in time \cite{Tang:2012, Gagarina:2013, Zhao:2014, Campos:2014, Zhao:2014, Oberblobaum:2014, Gagarina:2016, Muehlebach:2016, Ober-Blobaum:2016}.

Our low-order scheme is obtained by the algebraic up-winding of the high-order scheme as described in \cite{Kuzmin:2002, Kuzmin:2003, Moller:2013, Kuzmin:2021}. The result is a scheme similar to the well-known Godunov scheme. To integrate the low-order scheme in time we used the backward Euler method since the forward Euler method turned out to be unstable for certain cases in the neighbourhood of the center of the bubble $r=0$ and we exchanged the speed of the simulations in favour of robustness of results. The sacrifice is not very severe, since the up-winded advection matrix is band-limited with non-zero terms above diagonal only for nodes with negative velocity $v$ of plasma flow, so the implicit scheme can be efficiently implemented using Thomas' algorithm. 

The problem proved to be demanding and we had to develop a new limiting procedure that would work properly throughout the simulation. Our attempt is based on well-known Zalesak's limiter \cite{Zalesak:1979} in its peak-preserving version \cite{Zalesak:2012} corrected by the idea inspired by \cite{Kunhardt:1987} to restrict distances from which the conserved quantity values should be considered. The main observation is that the time step $\delta t$ used to integrate the equations in time needs to satisfy the Courant-Friedrichs-Lewy condition, i.e.
\begin{equation}
    \delta t  \le C \min{\left(\frac{\delta r}{|v|}\right)},
\end{equation}
where $C$ is the so-called Courant number which depends on the discretization scheme and $|v|$ is the maximal speed of propagation. For explicit time integration typically $C \lesssim 1$, thus the distance from which the conserved quantity can be transported in a time step to a node is bounded by the product $\delta t |v|$ which must be smaller than the lattice spacing $\delta r$. As a result, it is more consistent to use values of the conserved quantity in a distance of $\delta t |v|$ in the limiter instead of values from the neighbouring node directly. Even though, we conservatively assumed that the maximal speed is the speed of light, this correction significantly improved robustness of our scheme.

Finally, the right-hand side of equations \eqref{eq:nonlinear_wave_equation} and \eqref{eq:plasma_equations_of_motion} can be consistently discretized using the Galerkin method with interpolation functions introduced above. Even though, interpolation functions in time were chosen in such a way to obtain explicit schemes even when $\eta$ dependent terms are included, the implicit term for $\dot{\phi}$ arises from the right-hand side of \eqref{eq:nonlinear_wave_equation}. Fortunately, the dependence on $\dot{\phi}$ is linear and the implicit equation can be solved exactly.

\section{Benchmark points}
In order to evaluate the performance of our code, we initialized runs with two benchmark points that were already studied in a similar context~\cite{Hindmarsh:2013xza}. Table~\ref{tab:models} summarises all important parameters characterising these models. 

\begin{table}[b]%The best place to locate the table environment is directly after its first reference in text
\begin{ruledtabular}
\begin{tabular}{cccccccc}
\textrm{Model}&
\textrm{$T_c$ [GeV]}&
\textrm{$T_0$ [GeV]}&
\textrm{$\gamma$}&
\textrm{$\delta$}&
\textrm{$\lambda$}&
\textrm{$T_n$ [GeV]}&
\textrm{$\alpha_{\theta}$}\\

\colrule
$M_1$ & $100$ & $\frac{100}{\sqrt{2}}$ & $\frac{1}{18}$ & $\frac{\sqrt{10}}{72}$ & $\frac{10}{648}$ & $86$ & 0.005\\
         $M_2$ & $100$ & $\frac{100}{\sqrt{2}}$ & $\frac{2}{18}$ & $\frac{\sqrt{10}}{72}$ & $\frac{5}{648}$  & $80$ & 0.05 \\
\end{tabular}
\end{ruledtabular}
\caption{
Model parameters for the benchmark points
}
\label{tab:models}
\end{table}

For both of them, we perform a scan with respect to the friction parameter $\eta$, which is the only free parameter.
In accordance with previous studies, steady-state wall velocity $\xi_w$ grows as the friction becomes smaller~\cite{Kurki-Suonio:1996wfr}. The general shape of this correspondence was also confirmed, however, the exact form of the curve depends on the choice of the potential parameters as discussed in the main text.

% \begin{figure}[!t]
% \centering
% \includegraphics[scale=1]{summary_M1.pdf}
% \includegraphics[scale=1]{summary_profiles_M1.pdf}
% \caption{\it Weaker transition. Relation between the friction parameter $\eta$ and bubble-wall velocity in the stationary state $\xi_w$ (upper panel). Overview of the corresponding velocity profiles for each point (lower panel)}
% \label{fig:M1}
% \end{figure}

% \begin{figure}[!t]
% \centering
% \includegraphics[scale=1]{summary_M2.pdf}
% \includegraphics[scale=1]{summary_profiles_M2.pdf}
% \caption{\it Stronger transition. Relation between the friction parameter $\eta$ and bubble-wall velocity in the stationary state $\xi_w$ (upper panel). Overview of the corresponding velocity profiles for each point (lower panel)}
% \label{fig:M2}
% \end{figure}

\begin{figure*}[!t]
\centering
\includegraphics[scale=1]{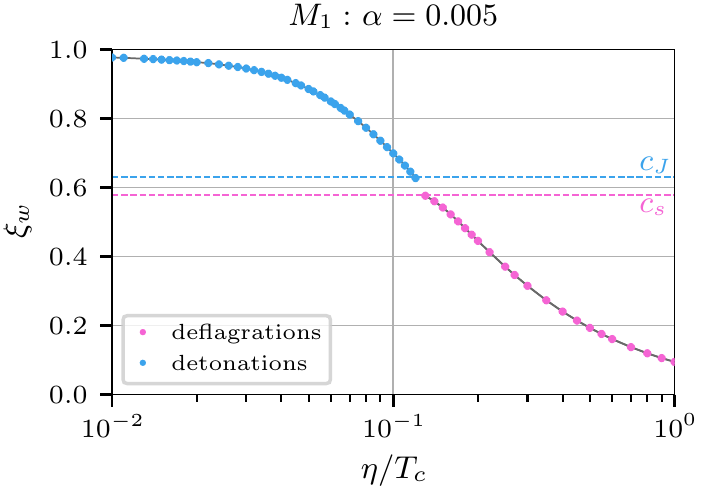}
\includegraphics[scale=1]{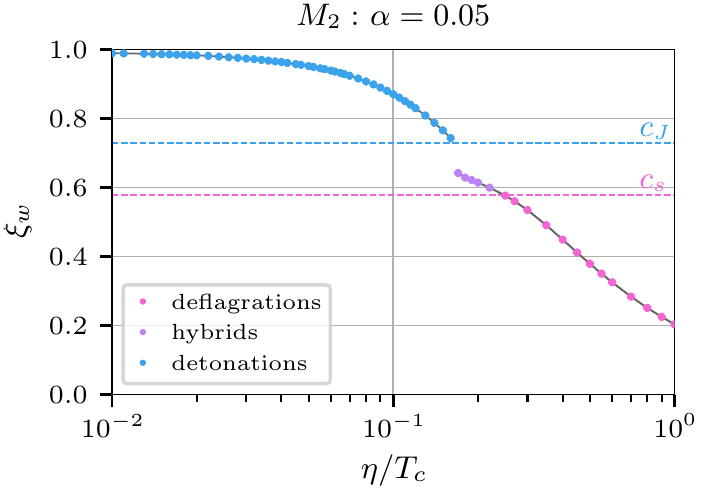}
\includegraphics[scale=1]{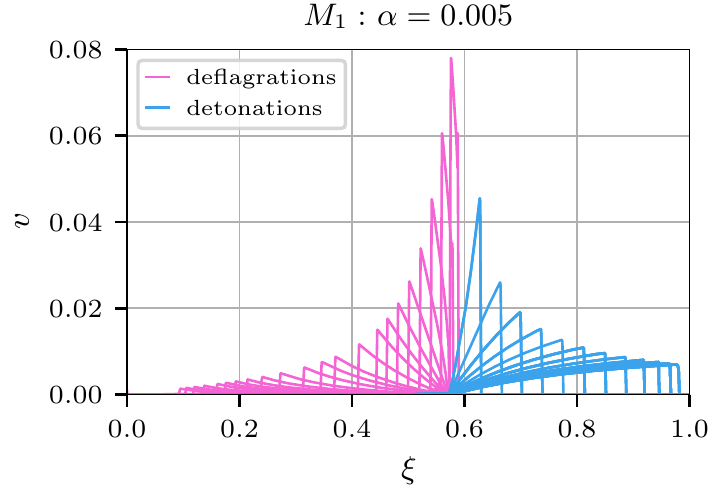}
\includegraphics[scale=1]{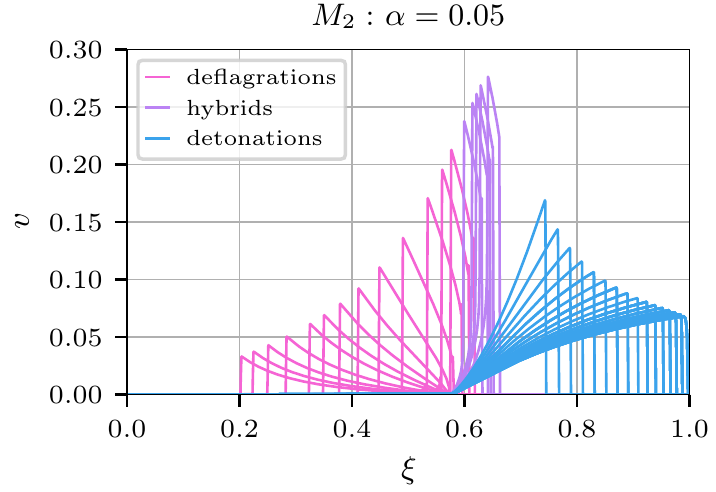}
\caption{\it 
Relation between the friction parameter $\eta$ and bubble-wall velocity in the stationary state $\xi_w$ (upper row). Overview of the corresponding velocity profiles for each point (lower row). The left column corresponds to the weaker transition ($M_1$), while the right one to the stronger one ($M_2$).}
\label{fig:M1M2}
\end{figure*}

Results of the scan are presented in Fig \ref{fig:M1M2}. We managed to obtain all three types of solutions for the stronger transition $(M_2)$ while for the weaker one $(M_1)$ no hybrids were realised. We, therefore, confirm the presence of the velocity gap in the region where one expects hybrid profiles. Such a gap appears in both cases, covering the whole $\xi_w\in[c_s, c_J]$ range for $M_1$ and allowing to continue deflagration branch towards solutions with supersonic wall velocity for $M_2$. The details of this phenomenon were not well understood so far and constitute the focus of our interest in the main text. 

\begin{figure*}[ht]
\centering
\includegraphics[scale=1]{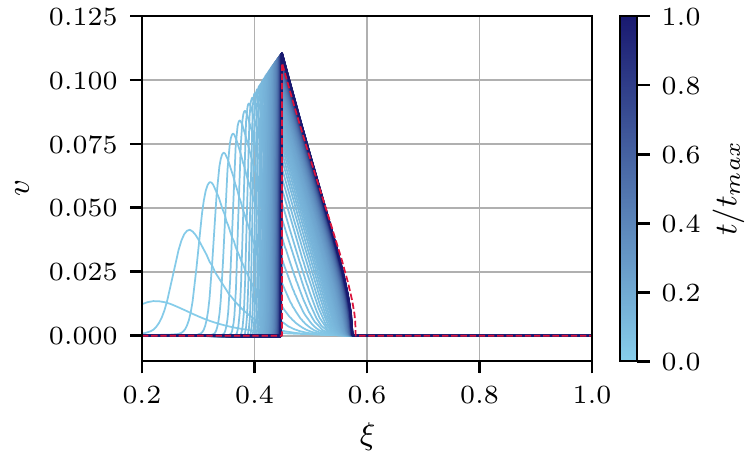}
\includegraphics[scale=1]{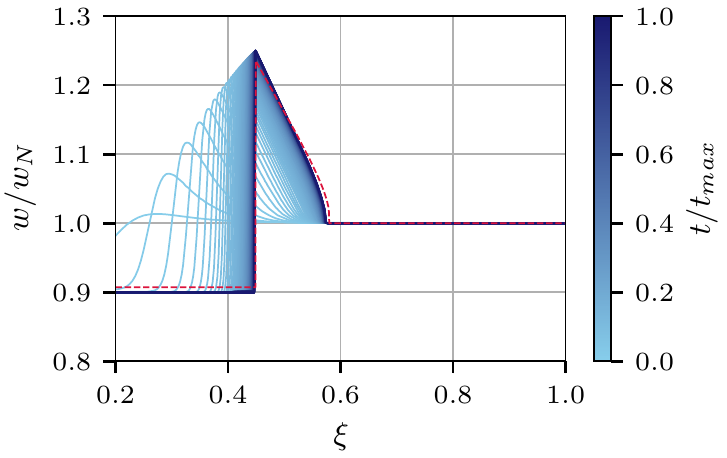}
\includegraphics[scale=1]{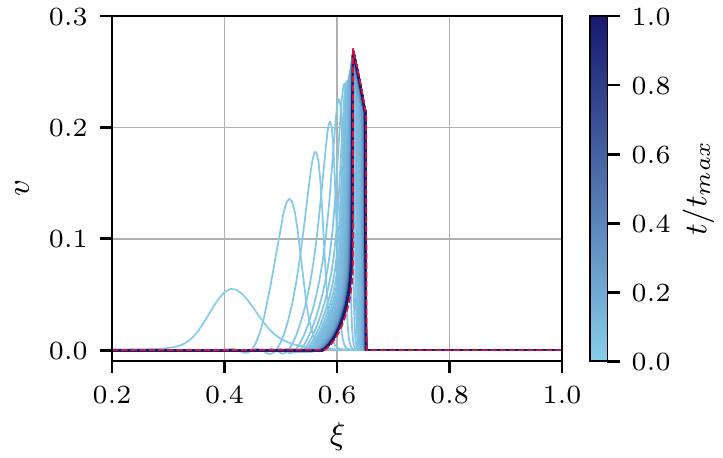}
\includegraphics[scale=1]{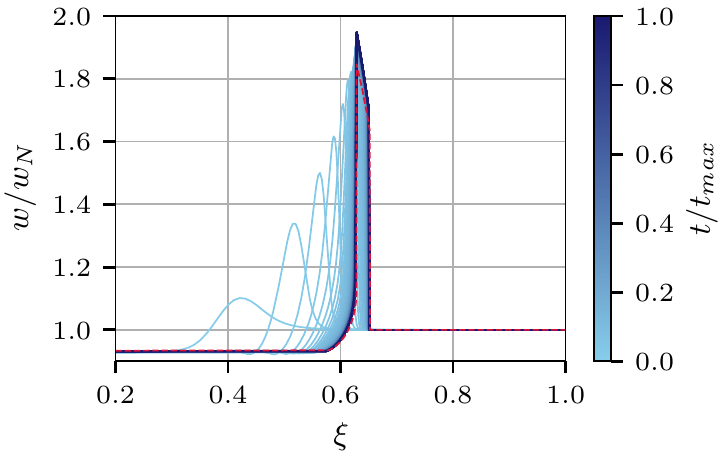}
\includegraphics[scale=1]{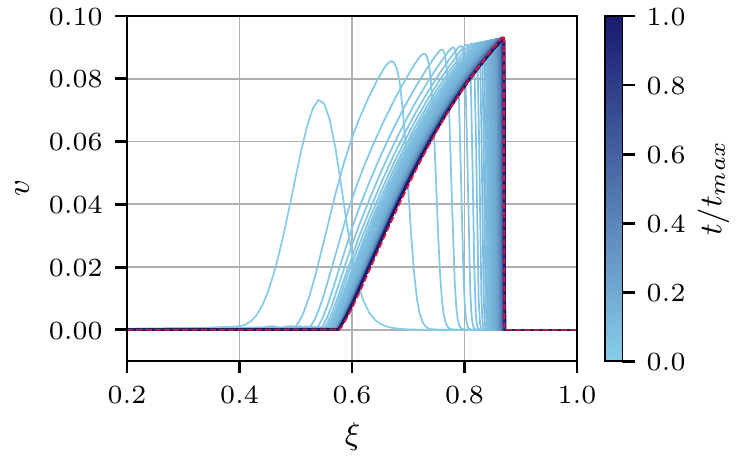}
\includegraphics[scale=1]{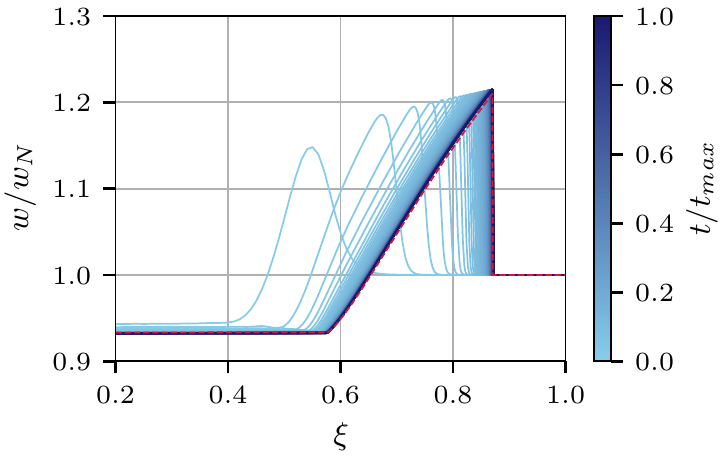}
\caption{\it Time evolution of the plasma shell profiles for plasma velocity $v$ (left column) and enthalpy $w$ (right column) as functions of the self-similar variable $\xi=r/t$. In general, all three types of solutions were found: deflagrations (first row), hybrids (second row) and detonations (third row). Different shades of blue correspond with the time flow. Profiles evolve towards stationary states represented by darker colours. Red, dashed lines denote analytical profiles obtained within the bag model.} 
\label{fig:comparison}
\end{figure*}

The shapes of stationary profiles are in very good agreement with the predictions of the bag model~\footnote{Recent $N-$body simulations~\cite{Lewicki:2022nba} foregoing perfect fluid and treating plasma as individual particles also found qualitatively the same fluid solutions.}. The comparison of the results of our simulations and the analytical profiles for three representative examples is shown in Fig.~\ref{fig:comparison}. As one can see, we managed to resolve the shocks and reproduce the form of hybrids with very good accuracy, which typically was challenging in existing results involving dynamical codes.

%%%%%%%%%%%%%%%%%%%%%%%%%%%%%%%%%%%%%%%%%%%%%%%%%%%%%%%%%%%%%%%%%%%
%%%%%%%%%%%%%%%%%%%%%%%%%%%%%%%%%%%%%%%%%%%%%%%%%%%%%%%%%%%%%%%%%%%
%%%%%%%%%%%%%%%%%%%%%%%%%%%%%%%%%%%%%%%%%%%%%%%%%%%%%%%%%%%%%%%%%%%
%%%%%%%%%%%%%%%%%%%%%%%%%%%%%%%%%%%%%%%%%%%%%%%%%%%%%%%%%%%%%%%%%%%
%%%%%%%%%%%%%%%%%%%%%%%%%%%%%%%%%%%%%%%%%%%%%%%%%%%%%%%%%%%%%%%%%%%
%%%%%%%%%%%%%%%%%%%%%%%%%%%%%%%%%%%%%%%%%%%%%%%%%%%%%%%%%%%%%%%%%%%
%%%%%%%%%%%%%%%%%%%%%%%%%%%%%%%%%%%%%%%%%%%%%%%%%%%%%%%%%%%%%%%%%%%
\section{Simplified transition analysis for polynomial potentials}\label{sec:poly_potentials}

Probability of tunneling at temperature $T$ is computed from the bubble nucleation rate~\cite{Coleman:1977py,Callan:1977pt,Linde:1980tt,Linde:1981zj}
\begin{equation}
\Gamma(T) = A(T)\textrm{e}^{-S}\, .
\end{equation}
For tunneling in finite temperatures the Euclidean action $S=\frac{S_3}{T}$ and 
$A(T)=T^4\left(\frac{S_3}{2\pi T}\right)^{\frac{3}{2}}$.
In order to obtain the critical bubble, one needs to find the nucleation temperature $T_n$ at which the 
probability of a true vacuum bubble forming within a horizon radius 
becomes significant~\cite{Ellis:2018mja} \, 
\begin{equation}
N(T_n) = \int_{T_n}^{T_c} \frac{dT}{T} \frac{\Gamma(T)}{H(T)^4} \approx 1, 
\end{equation}
where $T_c$ denotes the critical temperature in which both minima are degenerate. Assuming $H(t)\approx$ const, this condition reduces to
\begin{equation}
\frac{S_3}{T_n}\approx 4\log\left(\frac{T_n}{H}\right),
\end{equation}
which for temperatures around the electroweak scale gives $S_3/T_n \approx 140$~\cite{Caprini:2019egz}.
The simplest polynomial renormalizable potential takes the form
\begin{equation}
    V(\phi) = m^2\phi^2 - a\phi^3 + \lambda \phi^4,
\end{equation}
where $m^2$, $a$ and $\lambda$ may depend on the temperature. For such potential, there exists an accurate semi-analytical approximation of the critical action\cite{Adams:1993zs, Ellis:2020awk}:
\begin{equation}
    \frac{S_3}{T} = \frac{a}{T\lambda^{3/2}}\frac{8\pi\sqrt{\beta_1\delta + \beta_2\delta^2 + \beta_3\delta^3}}{81(2-\delta)^2},
\end{equation}
where $\delta$ = $\frac{8\lambda m^2}{a^2}$ and $\beta_1 = 8.2938$, $\beta_2 = -5.5330$, $\beta_3 = 0.8180$.
Therefore the nucleation rate for the potential in question, namely
\begin{equation}
V(\phi,T) = \frac{1}{2}\gamma(T^2-T_0^2)\phi^2 - \frac{1}{3}\delta T\phi^3 + \frac{1}{4}\lambda\phi^4\, .
\end{equation}
may be estimated as
\begin{widetext}
\begin{equation}
    \frac{\Gamma}{H^4} = \frac{T^4\exp{(-\frac{S_3}{T}})}{\rho_r + \rho_V}=\frac{\exp{(-\frac{8\pi\alpha(\frac{4\delta}{\lambda})^{3/2}(\beta_1 + \beta_2\delta + \beta_3\delta^2)}{243(\delta-2)^2})}}{\left( \frac{T_0^2}{1-\frac{\pi^2\alpha^2}{9\pi^2 \gamma\lambda}} \delta\right)^2\left(\frac{1}{3M_{pl}^2}\right)^2\left(\frac{\pi^2 g_{*}}{30} + (4\pi\alpha)^4\frac{(\sqrt{9-4\delta}+3)^2(\sqrt{9-4\delta}+3-2\delta)}{2\cdot 24^4\pi^4\lambda^3}\right)^2}.
\end{equation}
\end{widetext}
This expression depends on the temperature only through dimensionless parameter $\delta$, which varies from $\delta=0$ at $T_0$ to $\delta=2$ at $T_c$. Using this significantly simplifies the calculations, as the value of $\delta$ for which $\frac{\Gamma}{H^4}=1$ can be easily translated into $T_n$.
%%%%%%%%%%%%%%%%%%%%%%%%%%%%%%%%%%%%%%%%%%%%%%%%%%%%%%%%%%%%%%%%%%%%%%%%%%%%%%%%%%%%%%%%%%%%%%%%%%%%%%%%%%%%
\section{Impact on specific model results}
%%%%%%%%%%%%%%%%%%%%%%%%%%%%%%%%%%%%%%%%%%%%%%%%%%%%%%%%%%%%%%%%%%%%%%%%%%%%%%%%%%%%%%%%%%%%%%%%%%%%%%%%%%%%%%

In order to assess the impact of the hydrodynamical obstruction observed in our simulations, we checked our criterion against the points from scans presented in ref.~\cite{Ellis:2022lft}. It dealt with electroweak baryogenesis in the Standard Model supplemented by a neutral singlet. In order to accurately compute the baryon yield it used WKB approximation to compute the wall velocity~\cite{Cline:2020jre, Cline:2021dkf, Lewicki:2021pgr, Laurent:2022jrs, Ellis:2022lft} using the improved expansion of the fluid perturbations~\cite{ Laurent:2020gpg}. However, this scheme still relied on the hydrodynamical solution for a self-similar wall found using the bag model~\cite{Espinosa:2010hh} which does not take into account the shape of the potential and simply predicts a solution for every chosen wall velocity.

\begin{figure}
    \centering
    \includegraphics[scale=0.5]{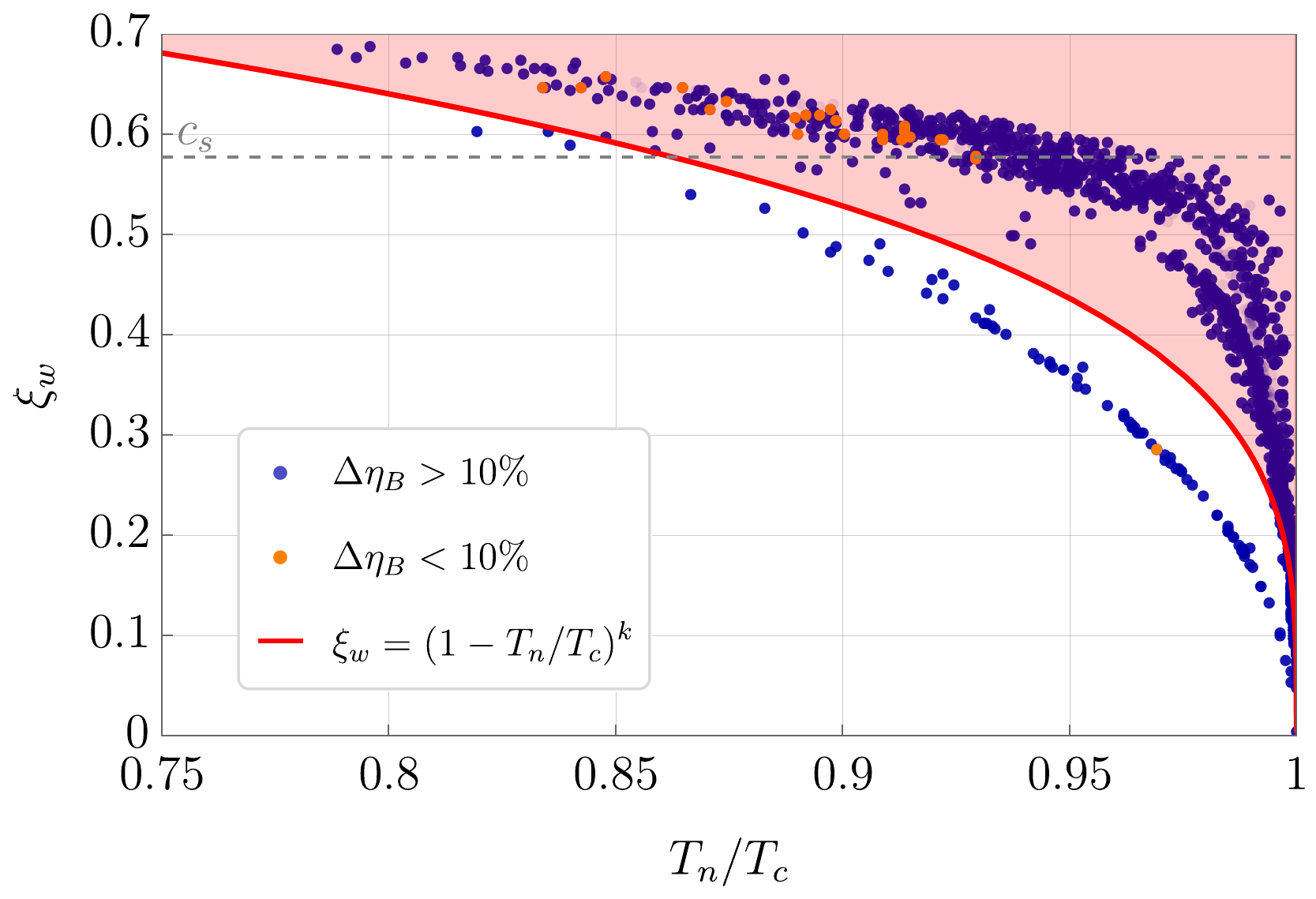}
    \caption{\it{Velocities of bubble walls in the specific model of scalar singlet extension \cite{Ellis:2022lft} evaluated using WKB approximation (colorful points) compared with the numerical fit derived in this paper (red curve). The most promising candidates for successfully baryogenesis i.e. the points with baryon to photon ratio $\eta_B$ consistent with the observed value up to $10\%$ are denoted in orange, while the rest is denoted in blue. Almost all of them lay in the region that might be impacted (shaded area). The dashed, gray line represents the speed of sound $c_s$.}}
    \label{fig:comparison-singlet}
\end{figure}

We now improve this by checking against our fit from the main text whether the assumed solutions can truly be realised in the potential of that model. The results are shown in Fig.~\ref{fig:comparison-singlet}.
We find that nearly all wall velocities predicted there will be lowered as the walls cease to accelerate due to the obstruction before achieving the velocities predicted by the WKB approximation. Moreover, the most promising candidates for successful baryogenesis are typically the fastest hybrids (orange points) and we see also all of them will be impacted. 
%Fig.~\ref{fig:comparison-singlet} clearly shows that the effect discussed here should be taken into account as the additional constraint on the velocity of the bubble wall. As direct verification of this mechanism in particle physics models is beyond the scope of this work, the precise $\alpha$-dependent evaluations are left for future. 
%Evaluation of the exact constraints and testing compliance with the calculations of the WKB approximation would require a simulation for each of the points to be verified, which should be performed carefully espacially in the limit $T_n/Tc\to 1$, where the transitions are typically the weakest in the realistic models. However direct verification of this mechanism in particle physics models is beyond the scope of this work and is left for the future.
Evaluation of the exact constraints would in principle require a simulation for each of the points to be verified. Especially in the limit $T_n/T_c\to 1$, where the transitions are typically weaker than the benchmarks used in our simulations so far. However, direct verification of this mechanism in particular particle physics models is beyond the scope of this paper and we leave it for future work.

\bibliography{main} 

\begin{thebibliography}{54}%
\makeatletter
\providecommand \@ifxundefined [1]{%
 \@ifx{#1\undefined}
}%
\providecommand \@ifnum [1]{%
 \ifnum #1\expandafter \@firstoftwo
 \else \expandafter \@secondoftwo
 \fi
}%
\providecommand \@ifx [1]{%
 \ifx #1\expandafter \@firstoftwo
 \else \expandafter \@secondoftwo
 \fi
}%
\providecommand \natexlab [1]{#1}%
\providecommand \enquote  [1]{``#1''}%
\providecommand \bibnamefont  [1]{#1}%
\providecommand \bibfnamefont [1]{#1}%
\providecommand \citenamefont [1]{#1}%
\providecommand \href@noop [0]{\@secondoftwo}%
\providecommand \href [0]{\begingroup \@sanitize@url \@href}%
\providecommand \@href[1]{\@@startlink{#1}\@@href}%
\providecommand \@@href[1]{\endgroup#1\@@endlink}%
\providecommand \@sanitize@url [0]{\catcode `\\12\catcode `\$12\catcode
  `\&12\catcode `\#12\catcode `\^12\catcode `\_12\catcode `\%12\relax}%
\providecommand \@@startlink[1]{}%
\providecommand \@@endlink[0]{}%
\providecommand \url  [0]{\begingroup\@sanitize@url \@url }%
\providecommand \@url [1]{\endgroup\@href {#1}{\urlprefix }}%
\providecommand \urlprefix  [0]{URL }%
\providecommand \Eprint [0]{\href }%
\providecommand \doibase [0]{http://dx.doi.org/}%
\providecommand \selectlanguage [0]{\@gobble}%
\providecommand \bibinfo  [0]{\@secondoftwo}%
\providecommand \bibfield  [0]{\@secondoftwo}%
\providecommand \translation [1]{[#1]}%
\providecommand \BibitemOpen [0]{}%
\providecommand \bibitemStop [0]{}%
\providecommand \bibitemNoStop [0]{.\EOS\space}%
\providecommand \EOS [0]{\spacefactor3000\relax}%
\providecommand \BibitemShut  [1]{\csname bibitem#1\endcsname}%
\let\auto@bib@innerbib\@empty
%</preamble>
\bibitem [{\citenamefont {Kuzmin}\ \emph {et~al.}(1985)\citenamefont {Kuzmin},
  \citenamefont {Rubakov},\ and\ \citenamefont {Shaposhnikov}}]{Kuzmin:1985mm}%
  \BibitemOpen
  \bibfield  {author} {\bibinfo {author} {\bibfnamefont {V.~A.}\ \bibnamefont
  {Kuzmin}}, \bibinfo {author} {\bibfnamefont {V.~A.}\ \bibnamefont {Rubakov}},
  \ and\ \bibinfo {author} {\bibfnamefont {M.~E.}\ \bibnamefont
  {Shaposhnikov}},\ }\href {\doibase 10.1016/0370-2693(85)91028-7} {\bibfield
  {journal} {\bibinfo  {journal} {Phys. Lett. B}\ }\textbf {\bibinfo {volume}
  {155}},\ \bibinfo {pages} {36} (\bibinfo {year} {1985})}\BibitemShut
  {NoStop}%
\bibitem [{\citenamefont {Cohen}\ \emph {et~al.}(1993)\citenamefont {Cohen},
  \citenamefont {Kaplan},\ and\ \citenamefont {Nelson}}]{Cohen:1993nk}%
  \BibitemOpen
  \bibfield  {author} {\bibinfo {author} {\bibfnamefont {A.~G.}\ \bibnamefont
  {Cohen}}, \bibinfo {author} {\bibfnamefont {D.~B.}\ \bibnamefont {Kaplan}}, \
  and\ \bibinfo {author} {\bibfnamefont {A.~E.}\ \bibnamefont {Nelson}},\
  }\href {\doibase 10.1146/annurev.ns.43.120193.000331} {\bibfield  {journal}
  {\bibinfo  {journal} {Ann. Rev. Nucl. Part. Sci.}\ }\textbf {\bibinfo
  {volume} {43}},\ \bibinfo {pages} {27} (\bibinfo {year} {1993})},\ \Eprint
  {http://arxiv.org/abs/hep-ph/9302210} {arXiv:hep-ph/9302210} \BibitemShut
  {NoStop}%
\bibitem [{\citenamefont {Rubakov}\ and\ \citenamefont
  {Shaposhnikov}(1996)}]{Rubakov:1996vz}%
  \BibitemOpen
  \bibfield  {author} {\bibinfo {author} {\bibfnamefont {V.~A.}\ \bibnamefont
  {Rubakov}}\ and\ \bibinfo {author} {\bibfnamefont {M.~E.}\ \bibnamefont
  {Shaposhnikov}},\ }\href {\doibase 10.1070/PU1996v039n05ABEH000145}
  {\bibfield  {journal} {\bibinfo  {journal} {Usp. Fiz. Nauk}\ }\textbf
  {\bibinfo {volume} {166}},\ \bibinfo {pages} {493} (\bibinfo {year}
  {1996})},\ \Eprint {http://arxiv.org/abs/hep-ph/9603208}
  {arXiv:hep-ph/9603208} \BibitemShut {NoStop}%
\bibitem [{\citenamefont {Morrissey}\ and\ \citenamefont
  {Ramsey-Musolf}(2012)}]{Morrissey:2012db}%
  \BibitemOpen
  \bibfield  {author} {\bibinfo {author} {\bibfnamefont {D.~E.}\ \bibnamefont
  {Morrissey}}\ and\ \bibinfo {author} {\bibfnamefont {M.~J.}\ \bibnamefont
  {Ramsey-Musolf}},\ }\href {\doibase 10.1088/1367-2630/14/12/125003}
  {\bibfield  {journal} {\bibinfo  {journal} {New J. Phys.}\ }\textbf {\bibinfo
  {volume} {14}},\ \bibinfo {pages} {125003} (\bibinfo {year} {2012})},\
  \Eprint {http://arxiv.org/abs/1206.2942} {arXiv:1206.2942 [hep-ph]}
  \BibitemShut {NoStop}%
\bibitem [{\citenamefont {Caprini}\ \emph {et~al.}(2016)\citenamefont {Caprini}
  \emph {et~al.}}]{Caprini:2015zlo}%
  \BibitemOpen
  \bibfield  {author} {\bibinfo {author} {\bibfnamefont {C.}~\bibnamefont
  {Caprini}} \emph {et~al.},\ }\href {\doibase 10.1088/1475-7516/2016/04/001}
  {\bibfield  {journal} {\bibinfo  {journal} {JCAP}\ }\textbf {\bibinfo
  {volume} {04}},\ \bibinfo {pages} {001} (\bibinfo {year} {2016})},\ \Eprint
  {http://arxiv.org/abs/1512.06239} {arXiv:1512.06239 [astro-ph.CO]}
  \BibitemShut {NoStop}%
\bibitem [{\citenamefont {Caprini}\ \emph {et~al.}(2020)\citenamefont {Caprini}
  \emph {et~al.}}]{Caprini:2019egz}%
  \BibitemOpen
  \bibfield  {author} {\bibinfo {author} {\bibfnamefont {C.}~\bibnamefont
  {Caprini}} \emph {et~al.},\ }\href {\doibase 10.1088/1475-7516/2020/03/024}
  {\bibfield  {journal} {\bibinfo  {journal} {JCAP}\ }\textbf {\bibinfo
  {volume} {03}},\ \bibinfo {pages} {024} (\bibinfo {year} {2020})},\ \Eprint
  {http://arxiv.org/abs/1910.13125} {arXiv:1910.13125 [astro-ph.CO]}
  \BibitemShut {NoStop}%
\bibitem [{\citenamefont {El-Neaj}\ \emph {et~al.}(2020)\citenamefont {El-Neaj}
  \emph {et~al.}}]{AEDGE:2019nxb}%
  \BibitemOpen
  \bibfield  {author} {\bibinfo {author} {\bibfnamefont {Y.~A.}\ \bibnamefont
  {El-Neaj}} \emph {et~al.} (\bibinfo {collaboration} {AEDGE}),\ }\href
  {\doibase 10.1140/epjqt/s40507-020-0080-0} {\bibfield  {journal} {\bibinfo
  {journal} {EPJ Quant. Technol.}\ }\textbf {\bibinfo {volume} {7}},\ \bibinfo
  {pages} {6} (\bibinfo {year} {2020})},\ \Eprint
  {http://arxiv.org/abs/1908.00802} {arXiv:1908.00802 [gr-qc]} \BibitemShut
  {NoStop}%
\bibitem [{\citenamefont {Badurina}\ \emph {et~al.}(2021)\citenamefont
  {Badurina}, \citenamefont {Buchmueller}, \citenamefont {Ellis}, \citenamefont
  {Lewicki}, \citenamefont {McCabe},\ and\ \citenamefont
  {Vaskonen}}]{Badurina:2021rgt}%
  \BibitemOpen
  \bibfield  {author} {\bibinfo {author} {\bibfnamefont {L.}~\bibnamefont
  {Badurina}}, \bibinfo {author} {\bibfnamefont {O.}~\bibnamefont
  {Buchmueller}}, \bibinfo {author} {\bibfnamefont {J.}~\bibnamefont {Ellis}},
  \bibinfo {author} {\bibfnamefont {M.}~\bibnamefont {Lewicki}}, \bibinfo
  {author} {\bibfnamefont {C.}~\bibnamefont {McCabe}}, \ and\ \bibinfo {author}
  {\bibfnamefont {V.}~\bibnamefont {Vaskonen}},\ }\href {\doibase
  10.1098/rsta.2021.0060} {\bibfield  {journal} {\bibinfo  {journal} {Phil.
  Trans. A. Math. Phys. Eng. Sci.}\ }\textbf {\bibinfo {volume} {380}},\
  \bibinfo {pages} {20210060} (\bibinfo {year} {2021})},\ \Eprint
  {http://arxiv.org/abs/2108.02468} {arXiv:2108.02468 [gr-qc]} \BibitemShut
  {NoStop}%
\bibitem [{\citenamefont {Cline}\ and\ \citenamefont
  {Kainulainen}(2020)}]{Cline:2020jre}%
  \BibitemOpen
  \bibfield  {author} {\bibinfo {author} {\bibfnamefont {J.~M.}\ \bibnamefont
  {Cline}}\ and\ \bibinfo {author} {\bibfnamefont {K.}~\bibnamefont
  {Kainulainen}},\ }\href {\doibase 10.1103/PhysRevD.101.063525} {\bibfield
  {journal} {\bibinfo  {journal} {Phys. Rev. D}\ }\textbf {\bibinfo {volume}
  {101}},\ \bibinfo {pages} {063525} (\bibinfo {year} {2020})},\ \Eprint
  {http://arxiv.org/abs/2001.00568} {arXiv:2001.00568 [hep-ph]} \BibitemShut
  {NoStop}%
\bibitem [{\citenamefont {Laurent}\ and\ \citenamefont
  {Cline}(2020)}]{Laurent:2020gpg}%
  \BibitemOpen
  \bibfield  {author} {\bibinfo {author} {\bibfnamefont {B.}~\bibnamefont
  {Laurent}}\ and\ \bibinfo {author} {\bibfnamefont {J.~M.}\ \bibnamefont
  {Cline}},\ }\href {\doibase 10.1103/PhysRevD.102.063516} {\bibfield
  {journal} {\bibinfo  {journal} {Phys. Rev. D}\ }\textbf {\bibinfo {volume}
  {102}},\ \bibinfo {pages} {063516} (\bibinfo {year} {2020})},\ \Eprint
  {http://arxiv.org/abs/2007.10935} {arXiv:2007.10935 [hep-ph]} \BibitemShut
  {NoStop}%
\bibitem [{\citenamefont {Dorsch}\ \emph {et~al.}(2021)\citenamefont {Dorsch},
  \citenamefont {Huber},\ and\ \citenamefont {Konstandin}}]{Dorsch:2021ubz}%
  \BibitemOpen
  \bibfield  {author} {\bibinfo {author} {\bibfnamefont {G.~C.}\ \bibnamefont
  {Dorsch}}, \bibinfo {author} {\bibfnamefont {S.~J.}\ \bibnamefont {Huber}}, \
  and\ \bibinfo {author} {\bibfnamefont {T.}~\bibnamefont {Konstandin}},\
  }\href {\doibase 10.1088/1475-7516/2021/08/020} {\bibfield  {journal}
  {\bibinfo  {journal} {JCAP}\ }\textbf {\bibinfo {volume} {08}},\ \bibinfo
  {pages} {020} (\bibinfo {year} {2021})},\ \Eprint
  {http://arxiv.org/abs/2106.06547} {arXiv:2106.06547 [hep-ph]} \BibitemShut
  {NoStop}%
\bibitem [{\citenamefont {Dorsch}\ \emph {et~al.}(2022)\citenamefont {Dorsch},
  \citenamefont {Huber},\ and\ \citenamefont {Konstandin}}]{Dorsch:2021nje}%
  \BibitemOpen
  \bibfield  {author} {\bibinfo {author} {\bibfnamefont {G.~C.}\ \bibnamefont
  {Dorsch}}, \bibinfo {author} {\bibfnamefont {S.~J.}\ \bibnamefont {Huber}}, \
  and\ \bibinfo {author} {\bibfnamefont {T.}~\bibnamefont {Konstandin}},\
  }\href {\doibase 10.1088/1475-7516/2022/04/010} {\bibfield  {journal}
  {\bibinfo  {journal} {JCAP}\ }\textbf {\bibinfo {volume} {04}},\ \bibinfo
  {pages} {010} (\bibinfo {year} {2022})},\ \Eprint
  {http://arxiv.org/abs/2112.12548} {arXiv:2112.12548 [hep-ph]} \BibitemShut
  {NoStop}%
\bibitem [{\citenamefont {Cline}\ \emph {et~al.}(2021)\citenamefont {Cline},
  \citenamefont {Friedlander}, \citenamefont {He}, \citenamefont {Kainulainen},
  \citenamefont {Laurent},\ and\ \citenamefont {Tucker-Smith}}]{Cline:2021iff}%
  \BibitemOpen
  \bibfield  {author} {\bibinfo {author} {\bibfnamefont {J.~M.}\ \bibnamefont
  {Cline}}, \bibinfo {author} {\bibfnamefont {A.}~\bibnamefont {Friedlander}},
  \bibinfo {author} {\bibfnamefont {D.-M.}\ \bibnamefont {He}}, \bibinfo
  {author} {\bibfnamefont {K.}~\bibnamefont {Kainulainen}}, \bibinfo {author}
  {\bibfnamefont {B.}~\bibnamefont {Laurent}}, \ and\ \bibinfo {author}
  {\bibfnamefont {D.}~\bibnamefont {Tucker-Smith}},\ }\href {\doibase
  10.1103/PhysRevD.103.123529} {\bibfield  {journal} {\bibinfo  {journal}
  {Phys. Rev. D}\ }\textbf {\bibinfo {volume} {103}},\ \bibinfo {pages}
  {123529} (\bibinfo {year} {2021})},\ \Eprint
  {http://arxiv.org/abs/2102.12490} {arXiv:2102.12490 [hep-ph]} \BibitemShut
  {NoStop}%
\bibitem [{\citenamefont {Cline}\ and\ \citenamefont
  {Laurent}(2021)}]{Cline:2021dkf}%
  \BibitemOpen
  \bibfield  {author} {\bibinfo {author} {\bibfnamefont {J.~M.}\ \bibnamefont
  {Cline}}\ and\ \bibinfo {author} {\bibfnamefont {B.}~\bibnamefont
  {Laurent}},\ }\href {\doibase 10.1103/PhysRevD.104.083507} {\bibfield
  {journal} {\bibinfo  {journal} {Phys. Rev. D}\ }\textbf {\bibinfo {volume}
  {104}},\ \bibinfo {pages} {083507} (\bibinfo {year} {2021})},\ \Eprint
  {http://arxiv.org/abs/2108.04249} {arXiv:2108.04249 [hep-ph]} \BibitemShut
  {NoStop}%
\bibitem [{\citenamefont {Lewicki}\ \emph {et~al.}(2022)\citenamefont
  {Lewicki}, \citenamefont {Merchand},\ and\ \citenamefont
  {Zych}}]{Lewicki:2021pgr}%
  \BibitemOpen
  \bibfield  {author} {\bibinfo {author} {\bibfnamefont {M.}~\bibnamefont
  {Lewicki}}, \bibinfo {author} {\bibfnamefont {M.}~\bibnamefont {Merchand}}, \
  and\ \bibinfo {author} {\bibfnamefont {M.}~\bibnamefont {Zych}},\ }\href
  {\doibase 10.1007/JHEP02(2022)017} {\bibfield  {journal} {\bibinfo  {journal}
  {JHEP}\ }\textbf {\bibinfo {volume} {02}},\ \bibinfo {pages} {017} (\bibinfo
  {year} {2022})},\ \Eprint {http://arxiv.org/abs/2111.02393} {arXiv:2111.02393
  [astro-ph.CO]} \BibitemShut {NoStop}%
\bibitem [{\citenamefont {Laurent}\ and\ \citenamefont
  {Cline}(2022)}]{Laurent:2022jrs}%
  \BibitemOpen
  \bibfield  {author} {\bibinfo {author} {\bibfnamefont {B.}~\bibnamefont
  {Laurent}}\ and\ \bibinfo {author} {\bibfnamefont {J.~M.}\ \bibnamefont
  {Cline}},\ }\href {\doibase 10.1103/PhysRevD.106.023501} {\bibfield
  {journal} {\bibinfo  {journal} {Phys. Rev. D}\ }\textbf {\bibinfo {volume}
  {106}},\ \bibinfo {pages} {023501} (\bibinfo {year} {2022})},\ \Eprint
  {http://arxiv.org/abs/2204.13120} {arXiv:2204.13120 [hep-ph]} \BibitemShut
  {NoStop}%
\bibitem [{\citenamefont {Ellis}\ \emph {et~al.}(2023)\citenamefont {Ellis},
  \citenamefont {Lewicki}, \citenamefont {Merchand}, \citenamefont {No},\ and\
  \citenamefont {Zych}}]{Ellis:2022lft}%
  \BibitemOpen
  \bibfield  {author} {\bibinfo {author} {\bibfnamefont {J.}~\bibnamefont
  {Ellis}}, \bibinfo {author} {\bibfnamefont {M.}~\bibnamefont {Lewicki}},
  \bibinfo {author} {\bibfnamefont {M.}~\bibnamefont {Merchand}}, \bibinfo
  {author} {\bibfnamefont {J.~M.}\ \bibnamefont {No}}, \ and\ \bibinfo {author}
  {\bibfnamefont {M.}~\bibnamefont {Zych}},\ }\href {\doibase
  10.1007/JHEP01(2023)093} {\bibfield  {journal} {\bibinfo  {journal} {JHEP}\
  }\textbf {\bibinfo {volume} {01}},\ \bibinfo {pages} {093} (\bibinfo {year}
  {2023})},\ \Eprint {http://arxiv.org/abs/2210.16305} {arXiv:2210.16305
  [hep-ph]} \BibitemShut {NoStop}%
\bibitem [{\citenamefont {No}(2011)}]{No:2011fi}%
  \BibitemOpen
  \bibfield  {author} {\bibinfo {author} {\bibfnamefont {J.~M.}\ \bibnamefont
  {No}},\ }\href {\doibase 10.1103/PhysRevD.84.124025} {\bibfield  {journal}
  {\bibinfo  {journal} {Phys. Rev. D}\ }\textbf {\bibinfo {volume} {84}},\
  \bibinfo {pages} {124025} (\bibinfo {year} {2011})},\ \Eprint
  {http://arxiv.org/abs/1103.2159} {arXiv:1103.2159 [hep-ph]} \BibitemShut
  {NoStop}%
\bibitem [{\citenamefont {Espinosa}\ \emph {et~al.}(2010)\citenamefont
  {Espinosa}, \citenamefont {Konstandin}, \citenamefont {No},\ and\
  \citenamefont {Servant}}]{Espinosa:2010hh}%
  \BibitemOpen
  \bibfield  {author} {\bibinfo {author} {\bibfnamefont {J.~R.}\ \bibnamefont
  {Espinosa}}, \bibinfo {author} {\bibfnamefont {T.}~\bibnamefont
  {Konstandin}}, \bibinfo {author} {\bibfnamefont {J.~M.}\ \bibnamefont {No}},
  \ and\ \bibinfo {author} {\bibfnamefont {G.}~\bibnamefont {Servant}},\ }\href
  {\doibase 10.1088/1475-7516/2010/06/028} {\bibfield  {journal} {\bibinfo
  {journal} {JCAP}\ }\textbf {\bibinfo {volume} {06}},\ \bibinfo {pages} {028}
  (\bibinfo {year} {2010})},\ \Eprint {http://arxiv.org/abs/1004.4187}
  {arXiv:1004.4187 [hep-ph]} \BibitemShut {NoStop}%
\bibitem [{\citenamefont {Konstandin}\ and\ \citenamefont
  {No}(2011)}]{Konstandin:2010dm}%
  \BibitemOpen
  \bibfield  {author} {\bibinfo {author} {\bibfnamefont {T.}~\bibnamefont
  {Konstandin}}\ and\ \bibinfo {author} {\bibfnamefont {J.~M.}\ \bibnamefont
  {No}},\ }\href {\doibase 10.1088/1475-7516/2011/02/008} {\bibfield  {journal}
  {\bibinfo  {journal} {JCAP}\ }\textbf {\bibinfo {volume} {02}},\ \bibinfo
  {pages} {008} (\bibinfo {year} {2011})},\ \Eprint
  {http://arxiv.org/abs/1011.3735} {arXiv:1011.3735 [hep-ph]} \BibitemShut
  {NoStop}%
\bibitem [{\citenamefont {Ignatius}\ \emph {et~al.}(1994)\citenamefont
  {Ignatius}, \citenamefont {Kajantie}, \citenamefont {Kurki-Suonio},\ and\
  \citenamefont {Laine}}]{Ignatius:1993qn}%
  \BibitemOpen
  \bibfield  {author} {\bibinfo {author} {\bibfnamefont {J.}~\bibnamefont
  {Ignatius}}, \bibinfo {author} {\bibfnamefont {K.}~\bibnamefont {Kajantie}},
  \bibinfo {author} {\bibfnamefont {H.}~\bibnamefont {Kurki-Suonio}}, \ and\
  \bibinfo {author} {\bibfnamefont {M.}~\bibnamefont {Laine}},\ }\href
  {\doibase 10.1103/PhysRevD.49.3854} {\bibfield  {journal} {\bibinfo
  {journal} {Phys. Rev. D}\ }\textbf {\bibinfo {volume} {49}},\ \bibinfo
  {pages} {3854} (\bibinfo {year} {1994})},\ \Eprint
  {http://arxiv.org/abs/astro-ph/9309059} {arXiv:astro-ph/9309059} \BibitemShut
  {NoStop}%
\bibitem [{\citenamefont {Kurki-Suonio}\ and\ \citenamefont
  {Laine}(1996)}]{Kurki-Suonio:1995yaf}%
  \BibitemOpen
  \bibfield  {author} {\bibinfo {author} {\bibfnamefont {H.}~\bibnamefont
  {Kurki-Suonio}}\ and\ \bibinfo {author} {\bibfnamefont {M.}~\bibnamefont
  {Laine}},\ }\href {\doibase 10.1103/PhysRevD.54.7163} {\bibfield  {journal}
  {\bibinfo  {journal} {Phys. Rev. D}\ }\textbf {\bibinfo {volume} {54}},\
  \bibinfo {pages} {7163} (\bibinfo {year} {1996})},\ \Eprint
  {http://arxiv.org/abs/hep-ph/9512202} {arXiv:hep-ph/9512202} \BibitemShut
  {NoStop}%
\bibitem [{\citenamefont {Kurki-Suonio}\ \emph {et~al.}(1997)\citenamefont
  {Kurki-Suonio}, \citenamefont {Jedamzik},\ and\ \citenamefont
  {Mathews}}]{Kurki-Suonio:1996wfr}%
  \BibitemOpen
  \bibfield  {author} {\bibinfo {author} {\bibfnamefont {H.}~\bibnamefont
  {Kurki-Suonio}}, \bibinfo {author} {\bibfnamefont {K.}~\bibnamefont
  {Jedamzik}}, \ and\ \bibinfo {author} {\bibfnamefont {G.~J.}\ \bibnamefont
  {Mathews}},\ }\href {\doibase 10.1086/303858} {\bibfield  {journal} {\bibinfo
   {journal} {Astrophys. J.}\ }\textbf {\bibinfo {volume} {479}},\ \bibinfo
  {pages} {31} (\bibinfo {year} {1997})},\ \Eprint
  {http://arxiv.org/abs/astro-ph/9606011} {arXiv:astro-ph/9606011} \BibitemShut
  {NoStop}%
\bibitem [{\citenamefont {Hindmarsh}\ \emph {et~al.}(2014)\citenamefont
  {Hindmarsh}, \citenamefont {Huber}, \citenamefont {Rummukainen},\ and\
  \citenamefont {Weir}}]{Hindmarsh:2013xza}%
  \BibitemOpen
  \bibfield  {author} {\bibinfo {author} {\bibfnamefont {M.}~\bibnamefont
  {Hindmarsh}}, \bibinfo {author} {\bibfnamefont {S.~J.}\ \bibnamefont
  {Huber}}, \bibinfo {author} {\bibfnamefont {K.}~\bibnamefont {Rummukainen}},
  \ and\ \bibinfo {author} {\bibfnamefont {D.~J.}\ \bibnamefont {Weir}},\
  }\href {\doibase 10.1103/PhysRevLett.112.041301} {\bibfield  {journal}
  {\bibinfo  {journal} {Phys. Rev. Lett.}\ }\textbf {\bibinfo {volume} {112}},\
  \bibinfo {pages} {041301} (\bibinfo {year} {2014})},\ \Eprint
  {http://arxiv.org/abs/1304.2433} {arXiv:1304.2433 [hep-ph]} \BibitemShut
  {NoStop}%
\bibitem [{\citenamefont {Hindmarsh}\ \emph {et~al.}(2015)\citenamefont
  {Hindmarsh}, \citenamefont {Huber}, \citenamefont {Rummukainen},\ and\
  \citenamefont {Weir}}]{Hindmarsh:2015qta}%
  \BibitemOpen
  \bibfield  {author} {\bibinfo {author} {\bibfnamefont {M.}~\bibnamefont
  {Hindmarsh}}, \bibinfo {author} {\bibfnamefont {S.~J.}\ \bibnamefont
  {Huber}}, \bibinfo {author} {\bibfnamefont {K.}~\bibnamefont {Rummukainen}},
  \ and\ \bibinfo {author} {\bibfnamefont {D.~J.}\ \bibnamefont {Weir}},\
  }\href {\doibase 10.1103/PhysRevD.92.123009} {\bibfield  {journal} {\bibinfo
  {journal} {Phys. Rev. D}\ }\textbf {\bibinfo {volume} {92}},\ \bibinfo
  {pages} {123009} (\bibinfo {year} {2015})},\ \Eprint
  {http://arxiv.org/abs/1504.03291} {arXiv:1504.03291 [astro-ph.CO]}
  \BibitemShut {NoStop}%
\bibitem [{Note1()}]{Note1}%
  \BibitemOpen
  \bibinfo {note} {Assuming local thermal equilibrium and neglecting the
  fluid-plasma coupling $\eta $ one can obtain analytical results~\cite
  {Balaji:2020yrx,Ai:2021kak,Ai:2023see} for the wall velocity.}\BibitemShut
  {Stop}%
\bibitem [{\citenamefont {Hindmarsh}\ \emph {et~al.}(2017)\citenamefont
  {Hindmarsh}, \citenamefont {Huber}, \citenamefont {Rummukainen},\ and\
  \citenamefont {Weir}}]{Hindmarsh:2017gnf}%
  \BibitemOpen
  \bibfield  {author} {\bibinfo {author} {\bibfnamefont {M.}~\bibnamefont
  {Hindmarsh}}, \bibinfo {author} {\bibfnamefont {S.~J.}\ \bibnamefont
  {Huber}}, \bibinfo {author} {\bibfnamefont {K.}~\bibnamefont {Rummukainen}},
  \ and\ \bibinfo {author} {\bibfnamefont {D.~J.}\ \bibnamefont {Weir}},\
  }\href {\doibase 10.1103/PhysRevD.96.103520} {\bibfield  {journal} {\bibinfo
  {journal} {Phys. Rev. D}\ }\textbf {\bibinfo {volume} {96}},\ \bibinfo
  {pages} {103520} (\bibinfo {year} {2017})},\ \bibinfo {note} {[Erratum:
  Phys.Rev.D 101, 089902 (2020)]},\ \Eprint {http://arxiv.org/abs/1704.05871}
  {arXiv:1704.05871 [astro-ph.CO]} \BibitemShut {NoStop}%
\bibitem [{\citenamefont {Auclair}\ \emph {et~al.}(2023)\citenamefont {Auclair}
  \emph {et~al.}}]{LISACosmologyWorkingGroup:2022jok}%
  \BibitemOpen
  \bibfield  {author} {\bibinfo {author} {\bibfnamefont {P.}~\bibnamefont
  {Auclair}} \emph {et~al.} (\bibinfo {collaboration} {LISA Cosmology Working
  Group}),\ }\href {\doibase 10.1007/s41114-023-00045-2} {\bibfield  {journal}
  {\bibinfo  {journal} {Living Rev. Rel.}\ }\textbf {\bibinfo {volume} {26}},\
  \bibinfo {pages} {5} (\bibinfo {year} {2023})},\ \Eprint
  {http://arxiv.org/abs/2204.05434} {arXiv:2204.05434 [astro-ph.CO]}
  \BibitemShut {NoStop}%
\bibitem [{\citenamefont {Hindmarsh}\ and\ \citenamefont
  {Hijazi}(2019)}]{Hindmarsh:2019phv}%
  \BibitemOpen
  \bibfield  {author} {\bibinfo {author} {\bibfnamefont {M.}~\bibnamefont
  {Hindmarsh}}\ and\ \bibinfo {author} {\bibfnamefont {M.}~\bibnamefont
  {Hijazi}},\ }\href {\doibase 10.1088/1475-7516/2019/12/062} {\bibfield
  {journal} {\bibinfo  {journal} {JCAP}\ }\textbf {\bibinfo {volume} {12}},\
  \bibinfo {pages} {062} (\bibinfo {year} {2019})},\ \Eprint
  {http://arxiv.org/abs/1909.10040} {arXiv:1909.10040 [astro-ph.CO]}
  \BibitemShut {NoStop}%
\bibitem [{\citenamefont {Ellis}\ \emph {et~al.}(2020)\citenamefont {Ellis},
  \citenamefont {Lewicki},\ and\ \citenamefont {No}}]{Ellis:2020awk}%
  \BibitemOpen
  \bibfield  {author} {\bibinfo {author} {\bibfnamefont {J.}~\bibnamefont
  {Ellis}}, \bibinfo {author} {\bibfnamefont {M.}~\bibnamefont {Lewicki}}, \
  and\ \bibinfo {author} {\bibfnamefont {J.~M.}\ \bibnamefont {No}},\ }\href
  {\doibase 10.1088/1475-7516/2020/07/050} {\bibfield  {journal} {\bibinfo
  {journal} {JCAP}\ }\textbf {\bibinfo {volume} {07}},\ \bibinfo {pages} {050}
  (\bibinfo {year} {2020})},\ \Eprint {http://arxiv.org/abs/2003.07360}
  {arXiv:2003.07360 [hep-ph]} \BibitemShut {NoStop}%
\bibitem [{\citenamefont {Steinhardt}(1982)}]{Steinhardt:1981ct}%
  \BibitemOpen
  \bibfield  {author} {\bibinfo {author} {\bibfnamefont {P.~J.}\ \bibnamefont
  {Steinhardt}},\ }\href {\doibase 10.1103/PhysRevD.25.2074} {\bibfield
  {journal} {\bibinfo  {journal} {Phys. Rev. D}\ }\textbf {\bibinfo {volume}
  {25}},\ \bibinfo {pages} {2074} (\bibinfo {year} {1982})}\BibitemShut
  {NoStop}%
\bibitem [{Note2()}]{Note2}%
  \BibitemOpen
  \bibinfo {note} {We neglect the possible deviations to spherical symmetry
  coming from instabilities that have been suggested in planar wall
  propagation~\cite {Kamionkowski:1992dc}.}\BibitemShut {Stop}%
\bibitem [{\citenamefont {Tang}\ and\ \citenamefont {Sun}(2012)}]{Tang:2012}%
  \BibitemOpen
  \bibfield  {author} {\bibinfo {author} {\bibfnamefont {W.}~\bibnamefont
  {Tang}}\ and\ \bibinfo {author} {\bibfnamefont {Y.}~\bibnamefont {Sun}},\
  }\href {\doibase 10.1016/j.amc.2012.08.062} {\bibfield  {journal} {\bibinfo
  {journal} {Applied Mathematics and Computation}\ }\textbf {\bibinfo {volume}
  {219}},\ \bibinfo {pages} {2158} (\bibinfo {year} {2012})}\BibitemShut
  {NoStop}%
\bibitem [{\citenamefont {Gagarina}\ \emph {et~al.}(2014)\citenamefont
  {Gagarina}, \citenamefont {Ambati}, \citenamefont {{van der Vegt}},\ and\
  \citenamefont {Bokhove}}]{Gagarina:2013}%
  \BibitemOpen
  \bibfield  {author} {\bibinfo {author} {\bibfnamefont {E.}~\bibnamefont
  {Gagarina}}, \bibinfo {author} {\bibfnamefont {V.}~\bibnamefont {Ambati}},
  \bibinfo {author} {\bibfnamefont {J.}~\bibnamefont {{van der Vegt}}}, \ and\
  \bibinfo {author} {\bibfnamefont {O.}~\bibnamefont {Bokhove}},\ }\href
  {\doibase https://doi.org/10.1016/j.jcp.2014.06.035} {\bibfield  {journal}
  {\bibinfo  {journal} {Journal of Computational Physics}\ }\textbf {\bibinfo
  {volume} {275}},\ \bibinfo {pages} {459} (\bibinfo {year}
  {2014})}\BibitemShut {NoStop}%
\bibitem [{\citenamefont {Zhao}\ and\ \citenamefont {Wei}(2014)}]{Zhao:2014}%
  \BibitemOpen
  \bibfield  {author} {\bibinfo {author} {\bibfnamefont {S.}~\bibnamefont
  {Zhao}}\ and\ \bibinfo {author} {\bibfnamefont {G.~W.}\ \bibnamefont {Wei}},\
  }\href {\doibase https://doi.org/10.1002/mma.2863} {\bibfield  {journal}
  {\bibinfo  {journal} {Mathematical Methods in the Applied Sciences}\ }\textbf
  {\bibinfo {volume} {37}},\ \bibinfo {pages} {1042} (\bibinfo {year}
  {2014})},\ \Eprint
  {http://arxiv.org/abs/https://onlinelibrary.wiley.com/doi/pdf/10.1002/mma.2863}
  {https://onlinelibrary.wiley.com/doi/pdf/10.1002/mma.2863} \BibitemShut
  {NoStop}%
\bibitem [{\citenamefont {Campos}(2014)}]{Campos:2014}%
  \BibitemOpen
  \bibfield  {author} {\bibinfo {author} {\bibfnamefont {C.~M.}\ \bibnamefont
  {Campos}},\ }\enquote {\bibinfo {title} {High order variational integrators:
  A polynomial approach},}\ in\ \href {\doibase 10.1007/978-3-319-06953-1_24}
  {\emph {\bibinfo {booktitle} {Advances in Differential Equations and
  Applications}}},\ \bibinfo {editor} {edited by\ \bibinfo {editor}
  {\bibfnamefont {F.}~\bibnamefont {Casas}}\ and\ \bibinfo {editor}
  {\bibfnamefont {V.}~\bibnamefont {Mart{\'i}nez}}}\ (\bibinfo  {publisher}
  {Springer International Publishing},\ \bibinfo {address} {Cham},\ \bibinfo
  {year} {2014})\ pp.\ \bibinfo {pages} {249--258}\BibitemShut {NoStop}%
\bibitem [{\citenamefont {Ober-Blöbaum}\ and\ \citenamefont
  {Saake}(2014)}]{Oberblobaum:2014}%
  \BibitemOpen
  \bibfield  {author} {\bibinfo {author} {\bibfnamefont {S.}~\bibnamefont
  {Ober-Blöbaum}}\ and\ \bibinfo {author} {\bibfnamefont {N.}~\bibnamefont
  {Saake}},\ }\href@noop {} {\enquote {\bibinfo {title} {Construction and
  analysis of higher order galerkin variational integrators},}\ } (\bibinfo
  {year} {2014}),\ \Eprint {http://arxiv.org/abs/1304.1398} {arXiv:1304.1398
  [math.NA]} \BibitemShut {NoStop}%
\bibitem [{\citenamefont {Gagarina}\ \emph {et~al.}(2016)\citenamefont
  {Gagarina}, \citenamefont {Ambati}, \citenamefont {Nurijanyan}, \citenamefont
  {{van der Vegt}},\ and\ \citenamefont {Bokhove}}]{Gagarina:2016}%
  \BibitemOpen
  \bibfield  {author} {\bibinfo {author} {\bibfnamefont {E.}~\bibnamefont
  {Gagarina}}, \bibinfo {author} {\bibfnamefont {V.}~\bibnamefont {Ambati}},
  \bibinfo {author} {\bibfnamefont {S.}~\bibnamefont {Nurijanyan}}, \bibinfo
  {author} {\bibfnamefont {J.}~\bibnamefont {{van der Vegt}}}, \ and\ \bibinfo
  {author} {\bibfnamefont {O.}~\bibnamefont {Bokhove}},\ }\href {\doibase
  https://doi.org/10.1016/j.jcp.2015.11.049} {\bibfield  {journal} {\bibinfo
  {journal} {Journal of Computational Physics}\ }\textbf {\bibinfo {volume}
  {306}},\ \bibinfo {pages} {370} (\bibinfo {year} {2016})}\BibitemShut
  {NoStop}%
\bibitem [{\citenamefont {Muehlebach}\ \emph {et~al.}(2016)\citenamefont
  {Muehlebach}, \citenamefont {Heimsch},\ and\ \citenamefont
  {Glocker}}]{Muehlebach:2016}%
  \BibitemOpen
  \bibfield  {author} {\bibinfo {author} {\bibfnamefont {M.}~\bibnamefont
  {Muehlebach}}, \bibinfo {author} {\bibfnamefont {T.}~\bibnamefont {Heimsch}},
  \ and\ \bibinfo {author} {\bibfnamefont {C.}~\bibnamefont {Glocker}}\
  }(\bibinfo {year} {2016})\BibitemShut {NoStop}%
\bibitem [{\citenamefont {Ober-Blöbaum}(2016)}]{Ober-Blobaum:2016}%
  \BibitemOpen
  \bibfield  {author} {\bibinfo {author} {\bibfnamefont {S.}~\bibnamefont
  {Ober-Blöbaum}},\ }\href {\doibase 10.1093/imanum/drv062} {\bibfield
  {journal} {\bibinfo  {journal} {IMA Journal of Numerical Analysis}\ }\textbf
  {\bibinfo {volume} {37}},\ \bibinfo {pages} {375} (\bibinfo {year} {2016})},\
  \Eprint
  {http://arxiv.org/abs/https://academic.oup.com/imajna/article-pdf/37/1/375/9633641/drv062.pdf}
  {https://academic.oup.com/imajna/article-pdf/37/1/375/9633641/drv062.pdf}
  \BibitemShut {NoStop}%
\bibitem [{\citenamefont {Kuzmin}\ and\ \citenamefont
  {Turek}(2002)}]{Kuzmin:2002}%
  \BibitemOpen
  \bibfield  {author} {\bibinfo {author} {\bibfnamefont {D.}~\bibnamefont
  {Kuzmin}}\ and\ \bibinfo {author} {\bibfnamefont {S.}~\bibnamefont {Turek}},\
  }\href {\doibase https://doi.org/10.1006/jcph.2001.6955} {\bibfield
  {journal} {\bibinfo  {journal} {Journal of Computational Physics}\ }\textbf
  {\bibinfo {volume} {175}},\ \bibinfo {pages} {525} (\bibinfo {year}
  {2002})}\BibitemShut {NoStop}%
\bibitem [{\citenamefont {Kuzmin}\ \emph {et~al.}(2003)\citenamefont {Kuzmin},
  \citenamefont {Möller},\ and\ \citenamefont {Turek}}]{Kuzmin:2003}%
  \BibitemOpen
  \bibfield  {author} {\bibinfo {author} {\bibfnamefont {D.}~\bibnamefont
  {Kuzmin}}, \bibinfo {author} {\bibfnamefont {M.}~\bibnamefont {Möller}}, \
  and\ \bibinfo {author} {\bibfnamefont {S.}~\bibnamefont {Turek}},\ }\href
  {\doibase https://doi.org/10.1002/fld.493} {\bibfield  {journal} {\bibinfo
  {journal} {International Journal for Numerical Methods in Fluids}\ }\textbf
  {\bibinfo {volume} {42}},\ \bibinfo {pages} {265} (\bibinfo {year} {2003})},\
  \Eprint
  {http://arxiv.org/abs/https://onlinelibrary.wiley.com/doi/pdf/10.1002/fld.493}
  {https://onlinelibrary.wiley.com/doi/pdf/10.1002/fld.493} \BibitemShut
  {NoStop}%
\bibitem [{\citenamefont {M\"{o}ller}(2013)}]{Moller:2013}%
  \BibitemOpen
  \bibfield  {author} {\bibinfo {author} {\bibfnamefont {M.}~\bibnamefont
  {M\"{o}ller}},\ }\href {\doibase 10.1007/s00607-012-0276-y} {\bibfield
  {journal} {\bibinfo  {journal} {Computing}\ }\textbf {\bibinfo {volume}
  {95}},\ \bibinfo {pages} {425–448} (\bibinfo {year} {2013})}\BibitemShut
  {NoStop}%
\bibitem [{\citenamefont {Kuzmin}(2021)}]{Kuzmin:2021}%
  \BibitemOpen
  \bibfield  {author} {\bibinfo {author} {\bibfnamefont {D.}~\bibnamefont
  {Kuzmin}},\ }\href {\doibase https://doi.org/10.1016/j.cma.2020.113569}
  {\bibfield  {journal} {\bibinfo  {journal} {Computer Methods in Applied
  Mechanics and Engineering}\ }\textbf {\bibinfo {volume} {373}},\ \bibinfo
  {pages} {113569} (\bibinfo {year} {2021})}\BibitemShut {NoStop}%
\bibitem [{\citenamefont {Zalesak}(1979)}]{Zalesak:1979}%
  \BibitemOpen
  \bibfield  {author} {\bibinfo {author} {\bibfnamefont {S.~T.}\ \bibnamefont
  {Zalesak}},\ }\href {\doibase https://doi.org/10.1016/0021-9991(79)90051-2}
  {\bibfield  {journal} {\bibinfo  {journal} {Journal of Computational
  Physics}\ }\textbf {\bibinfo {volume} {31}},\ \bibinfo {pages} {335}
  (\bibinfo {year} {1979})}\BibitemShut {NoStop}%
\bibitem [{\citenamefont {Zalesak}(2012)}]{Zalesak:2012}%
  \BibitemOpen
  \bibfield  {author} {\bibinfo {author} {\bibfnamefont {S.~T.}\ \bibnamefont
  {Zalesak}},\ }\enquote {\bibinfo {title} {The design of flux-corrected
  transport (fct) algorithms for structured grids},}\ in\ \href {\doibase
  10.1007/978-94-007-4038-9_2} {\emph {\bibinfo {booktitle} {Flux-Corrected
  Transport: Principles, Algorithms, and Applications}}},\ \bibinfo {editor}
  {edited by\ \bibinfo {editor} {\bibfnamefont {D.}~\bibnamefont {Kuzmin}},
  \bibinfo {editor} {\bibfnamefont {R.}~\bibnamefont {L{\"o}hner}}, \ and\
  \bibinfo {editor} {\bibfnamefont {S.}~\bibnamefont {Turek}}}\ (\bibinfo
  {publisher} {Springer Netherlands},\ \bibinfo {address} {Dordrecht},\
  \bibinfo {year} {2012})\ pp.\ \bibinfo {pages} {23--65}\BibitemShut {NoStop}%
\bibitem [{\citenamefont {Kunhardt}\ and\ \citenamefont
  {Wu}(1987)}]{Kunhardt:1987}%
  \BibitemOpen
  \bibfield  {author} {\bibinfo {author} {\bibfnamefont {E.}~\bibnamefont
  {Kunhardt}}\ and\ \bibinfo {author} {\bibfnamefont {C.}~\bibnamefont {Wu}},\
  }\href {\doibase https://doi.org/10.1016/0021-9991(87)90048-9} {\bibfield
  {journal} {\bibinfo  {journal} {Journal of Computational Physics}\ }\textbf
  {\bibinfo {volume} {68}},\ \bibinfo {pages} {127} (\bibinfo {year}
  {1987})}\BibitemShut {NoStop}%
\bibitem [{Note3()}]{Note3}%
  \BibitemOpen
  \bibinfo {note} {Recent $N-$body simulations~\cite {Lewicki:2022nba}
  foregoing perfect fluid and treating plasma as individual particles also
  found qualitatively the same fluid solutions.}\BibitemShut {Stop}%
\bibitem [{\citenamefont {Coleman}(1977)}]{Coleman:1977py}%
  \BibitemOpen
  \bibfield  {author} {\bibinfo {author} {\bibfnamefont {S.~R.}\ \bibnamefont
  {Coleman}},\ }\href {\doibase 10.1103/PhysRevD.16.1248} {\bibfield  {journal}
  {\bibinfo  {journal} {Phys. Rev. D}\ }\textbf {\bibinfo {volume} {15}},\
  \bibinfo {pages} {2929} (\bibinfo {year} {1977})},\ \bibinfo {note}
  {[Erratum: Phys.Rev.D 16, 1248 (1977)]}\BibitemShut {NoStop}%
\bibitem [{\citenamefont {Callan}\ and\ \citenamefont
  {Coleman}(1977)}]{Callan:1977pt}%
  \BibitemOpen
  \bibfield  {author} {\bibinfo {author} {\bibfnamefont {C.~G.}\ \bibnamefont
  {Callan}, \bibfnamefont {Jr.}}\ and\ \bibinfo {author} {\bibfnamefont
  {S.~R.}\ \bibnamefont {Coleman}},\ }\href {\doibase 10.1103/PhysRevD.16.1762}
  {\bibfield  {journal} {\bibinfo  {journal} {Phys. Rev. D}\ }\textbf {\bibinfo
  {volume} {16}},\ \bibinfo {pages} {1762} (\bibinfo {year}
  {1977})}\BibitemShut {NoStop}%
\bibitem [{\citenamefont {Linde}(1981)}]{Linde:1980tt}%
  \BibitemOpen
  \bibfield  {author} {\bibinfo {author} {\bibfnamefont {A.~D.}\ \bibnamefont
  {Linde}},\ }\href {\doibase 10.1016/0370-2693(81)90281-1} {\bibfield
  {journal} {\bibinfo  {journal} {Phys. Lett. B}\ }\textbf {\bibinfo {volume}
  {100}},\ \bibinfo {pages} {37} (\bibinfo {year} {1981})}\BibitemShut
  {NoStop}%
\bibitem [{\citenamefont {Linde}(1983)}]{Linde:1981zj}%
  \BibitemOpen
  \bibfield  {author} {\bibinfo {author} {\bibfnamefont {A.~D.}\ \bibnamefont
  {Linde}},\ }\href {\doibase 10.1016/0550-3213(83)90072-X} {\bibfield
  {journal} {\bibinfo  {journal} {Nucl. Phys. B}\ }\textbf {\bibinfo {volume}
  {216}},\ \bibinfo {pages} {421} (\bibinfo {year} {1983})},\ \bibinfo {note}
  {[Erratum: Nucl.Phys.B 223, 544 (1983)]}\BibitemShut {NoStop}%
\bibitem [{\citenamefont {Ellis}\ \emph {et~al.}(2019)\citenamefont {Ellis},
  \citenamefont {Lewicki},\ and\ \citenamefont {No}}]{Ellis:2018mja}%
  \BibitemOpen
  \bibfield  {author} {\bibinfo {author} {\bibfnamefont {J.}~\bibnamefont
  {Ellis}}, \bibinfo {author} {\bibfnamefont {M.}~\bibnamefont {Lewicki}}, \
  and\ \bibinfo {author} {\bibfnamefont {J.~M.}\ \bibnamefont {No}},\ }\href
  {\doibase 10.1088/1475-7516/2019/04/003} {\bibfield  {journal} {\bibinfo
  {journal} {JCAP}\ }\textbf {\bibinfo {volume} {04}},\ \bibinfo {pages} {003}
  (\bibinfo {year} {2019})},\ \Eprint {http://arxiv.org/abs/1809.08242}
  {arXiv:1809.08242 [hep-ph]} \BibitemShut {NoStop}%
\bibitem [{\citenamefont {Adams}(1993)}]{Adams:1993zs}%
  \BibitemOpen
  \bibfield  {author} {\bibinfo {author} {\bibfnamefont {F.~C.}\ \bibnamefont
  {Adams}},\ }\href {\doibase 10.1103/PhysRevD.48.2800} {\bibfield  {journal}
  {\bibinfo  {journal} {Phys. Rev. D}\ }\textbf {\bibinfo {volume} {48}},\
  \bibinfo {pages} {2800} (\bibinfo {year} {1993})},\ \Eprint
  {http://arxiv.org/abs/hep-ph/9302321} {arXiv:hep-ph/9302321} \BibitemShut
  {NoStop}%
\end{thebibliography}%
\end{document}